\documentclass[iop,apj,numberedappendix,onecolumn]{emulateapj}

\usepackage{natbib}
\usepackage{graphicx}
\usepackage{amsmath,amsfonts}
\usepackage{hyperref}
\usepackage{afterpage} 

\slugcomment{Accepted for publication in ApJ, April 9, 2015}

\def\sgra{Sgr~A$^{\ast}$}
\def\lsim{\mathrel{\raise.3ex\hbox{$<$\kern-.75em\lower1ex\hbox{$\sim$}}}}
\def\gsim{\mathrel{\raise.3ex\hbox{$>$\kern-.75em\lower1ex\hbox{$\sim$}}}}
\def\gtwid{\mathrel{\raise.3ex\hbox{$>$\kern-.75em\lower1ex\hbox{$\sim$}}}}
\def\proptwid{\mathrel{\raise.3ex\hbox{$\propto$\kern-.75em\lower1ex\hbox{$\sim$}}}}

\makeatletter
\newsavebox{\@brx}
\newcommand{\llangle}[1][]{\savebox{\@brx}{\(\m@th{#1\langle}\)}%
  \mathopen{\copy\@brx\kern-0.5\wd\@brx\usebox{\@brx}}}
\newcommand{\rrangle}[1][]{\savebox{\@brx}{\(\m@th{#1\rangle}\)}%
  \mathclose{\copy\@brx\kern-0.5\wd\@brx\usebox{\@brx}}}
\makeatother

\defcitealias{tms}{TMS}
\defcitealias{NarayanGoodman89}{NG89}
\defcitealias{GoodmanNarayan89}{GN89}
\setcitestyle{notesep={; }}

\begin{document}
\title{ Theory and Simulations of Refractive Substructure in Resolved Scatter-Broadened Images }
\shorttitle{Refractive Substructure}

\author{Michael D.\ Johnson\altaffilmark{1}, 
Carl R.\ Gwinn\altaffilmark{2}}
\shortauthors{Johnson \& Gwinn}
\altaffiltext{1}{Harvard-Smithsonian Center for Astrophysics, 60 Garden Street, Cambridge, MA 02138, USA}
\altaffiltext{2}{Department of Physics, University of California, Santa Barbara, CA 93106, USA}
\email{mjohnson@cfa.harvard.edu}

\begin{abstract}
At radio wavelengths, scattering in the interstellar medium distorts the appearance of astronomical sources. Averaged over a scattering ensemble, the result is a blurred image of the source. However, {\protect\NoHyper\citet{NarayanGoodman89}\protect\endNoHyper} and {\protect\NoHyper\citet{GoodmanNarayan89}\protect\endNoHyper} showed that for an incomplete average, scattering introduces refractive substructure in the image of a point source that is both persistent and wideband. We show that this substructure is quenched but not smoothed by an extended source. As a result, when the scatter-broadening is comparable to or exceeds the unscattered source size, the scattering can introduce spurious compact features into images. In addition, we derive efficient strategies to numerically compute realistic scattered images, and we present characteristic examples from simulations. Our results show that refractive substructure is an important consideration for ongoing missions at the highest angular 
resolutions, and we 
discuss specific 
implications for 
RadioAstron and the Event Horizon Telescope. 
\end{abstract}


\section{Introduction}
\label{sec::Introduction}

Radio-wave scattering in the turbulent interstellar medium (ISM) produces familiar effects: scintillation in frequency, time, and position. This scintillation has two distinct branches: diffractive and refractive. Diffractive scintillation is narrowband, short-lived, and is quenched by a source exceeding the diffractive scale, corresponding to the resolution of the scattering disk when viewed as a lens. As a result, diffractive scintillation is only observed for pulsars, masers, and a few extremely compact quasars. In contrast, refractive scintillation is wideband, persistent, but is quenched only when the angular size of the source exceeds that of the scattering disk. Reviews by \citet{Rickett_1990} and \citet{Narayan_1992} outline scattering theory and observations in both of these regimes.

\citet[][hereafter NG89]{NarayanGoodman89} and \citet[][hereafter GN89]{GoodmanNarayan89} uncovered a surprising refractive effect: substructure in the scattered image of a point source. Specifically, they showed that this substructure contributes noise to interferometric visibilities on baselines long enough to resolve the smooth, ensemble-average scattered image. This noise is wideband and persists over the refractive timescale (i.e., the time for the scattering material to move across the scattered image). \citet{Gwinn_2014} recently provided a dramatic demonstration of this effect through the discovery of substructure in the heavily scattered image of the Galactic center supermassive black hole, \sgra, at 1.3-cm wavelength.

We now extend the theoretical framework that describes refractive substructure to accommodate short interferometric baselines, anisotropic scattering, and extended source structure. We show that, surprisingly, substructure in the scattered image can occur at much finer angular scales than those of the unscattered source. Thus, although an extended source smoothes the diffraction pattern in the observing plane according to the familiar convolution action of scattering, it \emph{does not} smooth the scattered image but merely reduces the depth of fluctuations. As a result, refractive substructure can remain an important consideration for observations in which the scattering is somewhat subdominant to intrinsic structure, and it can introduce spurious compact features into resolved images of extended sources. 

We begin, in \S\ref{sec::Theoretical_Background} by reviewing some basic principles of scattering and scintillation. Next, in \S\ref{sec::Visibilities}, we explore interferometric visibilities in different averaging regimes. In \S\ref{sec::Scattered_Image}, we define the scattered image and consider how its appearance is affected by properties of the source and the scattering. We also derive expressions that allow for efficient numerical computation of scattered images and provide characteristic results from simulations. In \S\ref{sec::Manifestations_of_Refractive_Noise}, we derive specific observable properties of the refractive noise -- flux modulation, image wander, and substructure in scattered images -- using the second moment of the interferometric visibility modulus (derived in Appendix~\ref{sec::Refractive_Noise}). In \S\ref{sec::Observational_Implications}, we consider implications of refractive noise for two specific missions: RadioAstron and the Event Horizon Telescope. We summarize our findings 
in \S\ref{sec::Summary}.

\section{Theoretical Background}
\label{sec::Theoretical_Background}

\subsection{Interstellar Scattering and Scintillation}

Scattering of radio waves in the interstellar plasma arises from small-scale fluctuations in electron density. The resulting variations in refractive index produce variations in phase of the electromagnetic wave. As a result, each scattered ray has both geometrical and stochastic contributions to phase. 

In many cases, the scattering can be well-approximated by a single thin phase-changing screen. The change in phase by the screen is $\phi(\textbf{x})$, where $\textbf{x}$ is a transverse coordinate on the screen. The statistical characteristics of the scattering and scintillation can then be related to statistical characteristics of the phase fluctuations, either through a spatial structure function $D_{\phi}(\textbf{x}) \equiv \left \langle \left[ \phi\left(\textbf{x}_0 + \textbf{x} \right) - \phi\left(\textbf{x}_0 \right) \right]^2 \right \rangle$ or, equivalently, through the power spectrum $Q(\textbf{q})$ of phase fluctuations.

A variety of evidence \citep[e.g.,][]{Armstrong_1995} suggests that these functions are well approximated as power laws -- $D_{\phi}(\textbf{x}) \propto \left|\textbf{x} \right|^{\alpha}$ and $Q(\textbf{q}) \propto \left|\textbf{q} \right|^{-\left(\alpha+2\right)}$ -- spanning the immense range of scales from ${\sim} 10^2$\ km to ${>}1$\ AU.\footnote{Note that $\alpha$ is used inconsistently throughout the literature, with some authors choosing $\alpha$ as the index for $Q(\textbf{q})$ so that $D_{\phi}(\textbf{x}) \propto \left| \textbf{x} \right|^{\alpha-2}$.} On shorter scales, the phase fluctuations are smooth, so $D_{\phi}(\textbf{x}) \propto \left| \textbf{x} \right|^2$, while on longer scales, $D_{\phi}$ is constant \citep{Tatarskii_1971}. 

The density fluctuations, and their power-law form, may reflect the effects of a cascade of Alfv\'{e}n-wave turbulence \citep{Goldreich_1995,Lithwick_2001}.
Indeed, scattering often follows the Kolmogorov scaling expected for this and other types of turbulence, with power-law index $\alpha=5/3$.
In this picture, the cascade is initiated by driving forces at large spatial scales, the ``outer scale'' $r_{\rm out}$, and is terminated by dissipation at a minimum scale, the ``inner scale'' $r_{\rm in}$. Some evidence suggests that the inner scale for interstellar scattering material is 100 to 300~km \citep{Spangler_1990,Rickett_2009}, although only a few, highly-scattered lines of sight have been studied. The inner scale and turbulence properties may differ for paths along and perpendicular to the large-scale magnetic field. We provide simple expression for $D_{\phi}(\textbf{x})$ and $Q(\textbf{q})$ that accommodate anisotropy and inner and outer scales in Appendix \ref{sec::PSF} and \ref{sec::Refractive_Noise}. 

If the power-law index $\alpha < 2$ then the spectra is said to be ``shallow,'' whereas for $\alpha > 2$ the spectra is said to be ``steep.'' In this paper, we restrict our attention to shallow spectra.

\subsection{Scintillation and Averaging Regimes}
\label{sec::ScintillationAveragingRegimes}

In the strong scattering regime, there are three important length scales on the scattering screen. The \emph{phase coherence length}, $r_0$, corresponds to the separation between two points for which the root-mean-square (rms) screen phase difference is 1 radian: $D_{\phi}(r_0) \equiv 1$. The \emph{Fresnel scale}, $r_{\rm F}$, defines the lateral scale at which the change in geometrical phase relative to that of the direct path is 1/2 radian and is given explicitly in Eq.~\ref{eq::Fresnel} below. Finally, the \emph{refractive scale}, $r_{\rm R}$, determines the size of the scattered image of a point source:  $r_{\rm R} \equiv r_{\rm F}^2/r_0$. The strong scattering regime is defined by the condition $r_0 \ll r_{\rm F} \ll r_{\rm R}$. 

These length scales delineate the two scintillation regimes, diffractive and refractive, introduced in \S\ref{sec::Introduction}. Diffractive scintillation is dominated by fluctuations on the scale of $r_0$ whereas refractive scintillation is dominated by fluctuations on the scale of $r_{\rm R}$. Thus, diffractive scintillation decorrelates over a fractional bandwidth of ${\sim} r_0/r_{\rm R}$ while refractive scintillation decorrelates over a fractional bandwidth of unity. Likewise, if $v_{\perp}$ denotes the characteristic transverse velocity of the scattering material, then $r_0/v_{\perp}$ gives the diffractive timescale while $r_{\rm R}/v_{\perp}$ gives the refractive timescale. 

Following the treatment and nomenclature of \citetalias{GoodmanNarayan89}, we will consider three types of averages for quantities such as interferometric visibilities (denoted $V_x$) and images (denoted $I_x$). A \emph{snapshot} quantity (e.g., $V_{\rm ss}$) averages over source and background noise for a single realization of the scintillation pattern. An \emph{average} quantity (e.g., $V_{\rm a}$) averages also over diffractive scintillation but not refractive scintillation. An \emph{ensemble-average} quantity (e.g., $V_{\rm ea}$) averages over both diffractive and refractive scintillation. We will also use these subscripts to denote respective averages (e.g.\ $\langle \ldots \rangle_{\rm ss}$ represents a snapshot average).

\subsection{The Scalar Electric Field}

To derive observable consequences of scattering, we must first determine the scalar electric field $\psi(\textbf{b})$ at a transverse coordinate $\textbf{b}$ in the observing plane. Using the Fresnel diffraction integral, this field can be written as \citep[see, for instance,][]{ISO}
\begin{align}
\label{eq::field}
\psi(\textbf{b}) = \frac{1}{2\pi r_{\rm F}^2 } \int_{\rm screen} d^2 \textbf{x}\, e^{i \left[ \left(\frac{k}{2D} \right) |\textbf{b}-\textbf{x}|^2 + \phi(\textbf{x}) \right]} \int_{\rm src} d^2\textbf{s}\, e^{i \left(\frac{k}{2R} \right) |\textbf{x}-\textbf{s}|^2} \psi_{\rm src}(\textbf{s}).
\end{align}
In this expression, a large, constant phase is absorbed into the stochastic source field $\psi_{\rm src}(\textbf{s})$, and we have chosen the amplitude to simplify later calculations. 
Here, $\phi(\textbf{x})$ is the screen phase, $D$ is the characteristic Earth-scatterer distance, $R$ is the characteristic source-scatterer distance, $k=2\pi/\lambda$ is the wavenumber, and $r_{\rm F}$ is the Fresnel scale, defined by\footnote{Note that some authors define the Fresnel scale differently, often with the substitution $k \rightarrow k/(2\pi)$.} 
\begin{align}
\label{eq::Fresnel}
r_{\rm F} \equiv \sqrt{ \frac{ D R }{D + R} \frac{1}{k} }.
\end{align}
We will write most subsequent equations in terms of $r_{\rm F}$ and the effective magnification $M \equiv D/R$ of the scattering screen.

\section{Interferometric Visibilities in Different Averaging Regimes}
\label{sec::Visibilities}

We now consider the behavior of interferometric visibility (i.e., the electric field covariance) in the various averaging regimes outlined in \S\ref{sec::ScintillationAveragingRegimes}.

\subsection{The Snapshot Visibility}
\label{sec::Snapshot_Visibility}

\begin{figure}[t]
\centering
\includegraphics*[width=0.9\textwidth]{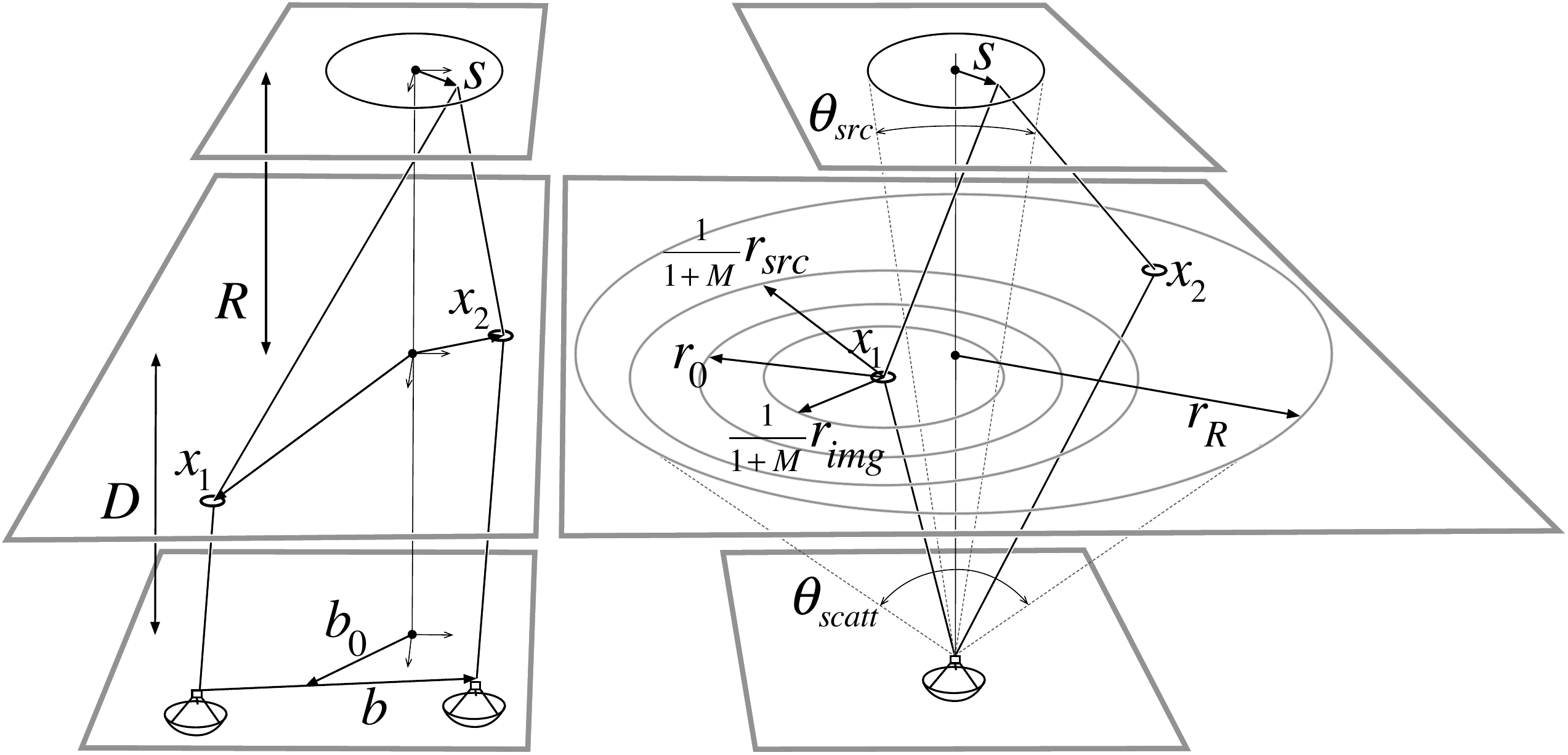}
\caption
{ 
Geometry relating the source, scattering, and observer. 
The scales $r_{\rm src}$ and $r_{\rm img}$ correspond to the coherence lengths in the observing plane of the electric field from the source and ensemble-average scattered images, respectively. These scales are related to the angular sizes of these images via $r_{\rm src} \sim 0.37 \lambda/\theta_{\rm src}$ and $r_{\rm img} \sim 0.37 \lambda/\theta_{\rm img}$, with $r_{\rm src} \geq r_{\rm img}$. Also, $r_0$ denotes the transverse scale on the scattering screen over which the rms phase difference is one radian; the scattered angular size of a point source is $\theta_{\rm scatt} \sim 0.37 \lambda/\left[(1+M)r_0 \right]$.
}
\label{fig_snapshot_visibility_geometry}
\end{figure}

The simplest averaging regime is the snapshot -- an average over noise of the source for a fixed realization of the scattering. We assume that the source is spatially incoherent with an intensity $I_{\rm src}(\textbf{s})$: $\left \langle \psi_{\rm src}(\textbf{s}) \psi_{\rm src}^\ast(\textbf{s}') \right \rangle_{\rm ss} \equiv I_{\rm src}(\textbf{s}) \delta\left( \textbf{s} - \textbf{s}' \right)$. Then, the snapshot visibility on a vector baseline $\textbf{b}$ centered on $\textbf{b}_0$ is (see Figure~\ref{fig_snapshot_visibility_geometry})
\begin{align}
\label{eq::Vsnapshot}
V_{\rm ss}(\textbf{b};\textbf{b}_0) &\equiv \left \langle \psi(\textbf{b}_0-\textbf{b}/2) \psi^\ast(\textbf{b}_0+\textbf{b}/2) \right \rangle_{\rm ss}\\
\nonumber &= \frac{1}{4\pi^2 r_{\rm F}^4} \int d^2 \textbf{x}_1\, d^2 \textbf{x}_2\, e^{i \frac{1}{2} r_{\rm F}^{-2} \left[ \left(x_1^2 - x_2^2\right) + \frac{\textbf{b}}{1+M} \cdot \left( \textbf{x}_1 + \textbf{x}_2 \right) - 2\frac{\textbf{b}_0}{1+M}\cdot\left(\textbf{b} + \textbf{x}_1 - \textbf{x}_2 \right) \right]} e^{i \left[ \phi(\textbf{x}_1) - \phi(\textbf{x}_2) \right]} \int d^2\textbf{s}\, e^{i \frac{k}{R} \left( \textbf{x}_2 - \textbf{x}_1 \right) \cdot \textbf{s}} I_{\rm src}(\textbf{s})\\
\nonumber &= \frac{1}{4\pi^2 r_{\rm F}^4} \int d^2 \textbf{x}_1\, d^2 \textbf{x}_2\, e^{i \frac{1}{2} r_{\rm F}^{-2} \left[ \left(x_1^2 - x_2^2\right) + \frac{\textbf{b}}{1+M} \cdot \left( \textbf{x}_1 + \textbf{x}_2 \right) - 2\frac{\textbf{b}_0}{1+M}\cdot\left(\textbf{b} + \textbf{x}_1 - \textbf{x}_2 \right) \right]} e^{i \left[ \phi(\textbf{x}_1) - \phi(\textbf{x}_2) \right]} V_{\rm src}(\left(1+M\right)\left(\textbf{x}_2 - \textbf{x}_1 \right)).
\end{align}
In the final expression, we have applied the Van Cittert-Zernike Theorem \citep[e.g.,][]{TMS} to replace the remaining source integral (a Fourier transform of the source brightness distribution) with its equivalent form as an interferometric visibility $V_{\rm src}$ corresponding to the \emph{unscattered} source image:
\begin{align}
V_{\rm src}(\textbf{b}) \propto \int d^2 \textbf{s}\,  I_{\rm src}(\textbf{s}) e^{2\pi i \left( \frac{\textbf{s}}{D+R} \right) \cdot \left( \frac{\textbf{b}}{\lambda} \right)}.
\end{align}
Because our later results are all expressed as fractional quantities, overall normalization is insignificant. To simplify calculations, we will adopt the normalization $V_{\rm src}(\textbf{0}) = 1$.

Note that the intensity of the diffraction (or illumination) pattern in the observing plane takes the form
\begin{align}
\label{eq::Isnapshot}
I_{\rm ss}(\textbf{b}_0) &\equiv \left \langle \psi(\textbf{b}_0) \psi^\ast(\textbf{b}_0) \right \rangle_{\rm ss}\\
\nonumber &= \frac{1}{4\pi^2 r_{\rm F}^4} \int d^2 \textbf{x}_1\, d^2 \textbf{x}_2\, e^{i \frac{1}{2} r_{\rm F}^{-2} \left[ \left(x_1^2 - x_2^2\right) - 2\frac{\textbf{b}_0}{1+M}\cdot\left(\textbf{x}_1 - \textbf{x}_2 \right) \right]} e^{i \left[ \phi(\textbf{x}_1) - \phi(\textbf{x}_2) \right]} \int d^2\textbf{s}\, e^{i \frac{k}{R} \left( \textbf{x}_2 - \textbf{x}_1 \right) \cdot \textbf{s}} I_{\rm src}(\textbf{s})\\
\nonumber &= \int d^2\textbf{s}\, I_{\rm src}(\textbf{s}) \times \left\{  \frac{1}{4\pi^2 r_{\rm F}^4} \int d^2 \textbf{x}_1\, d^2 \textbf{x}_2\, e^{i \frac{1}{2} r_{\rm F}^{-2} \left[ \left(x_1^2 - x_2^2\right)  \right]} e^{i \left[ \phi(\textbf{x}_1) - \phi(\textbf{x}_2) \right]} e^{i k \left( \frac{\textbf{s}}{R} + \frac{\textbf{b}_0}{D} \right) \left( \textbf{x}_2 - \textbf{x}_1 \right)} \right\}.
\end{align}
The term in braces is simply the point source diffraction pattern at a transverse coordinate $\textbf{b}_0 + M \textbf{s}$. So, for an extended source, we recover the well-known result that the diffraction pattern in the observing plane is the convolution of the point-source diffraction pattern with a magnified image of the source \citep{Salpeter_1967,Cohen_1967}. Hence, an extended source will smooth out any features in the diffraction pattern that are finer than the scales of the magnified source image. However, in \S\ref{sec::Scattered_Image} we will demonstrate that an extended source does \emph{not} smooth the scattered image.

\subsection{The Average Visibility}
\label{sec::A_Visibility}

\begin{figure}[t]
\centering
\includegraphics[width=0.6\textwidth]{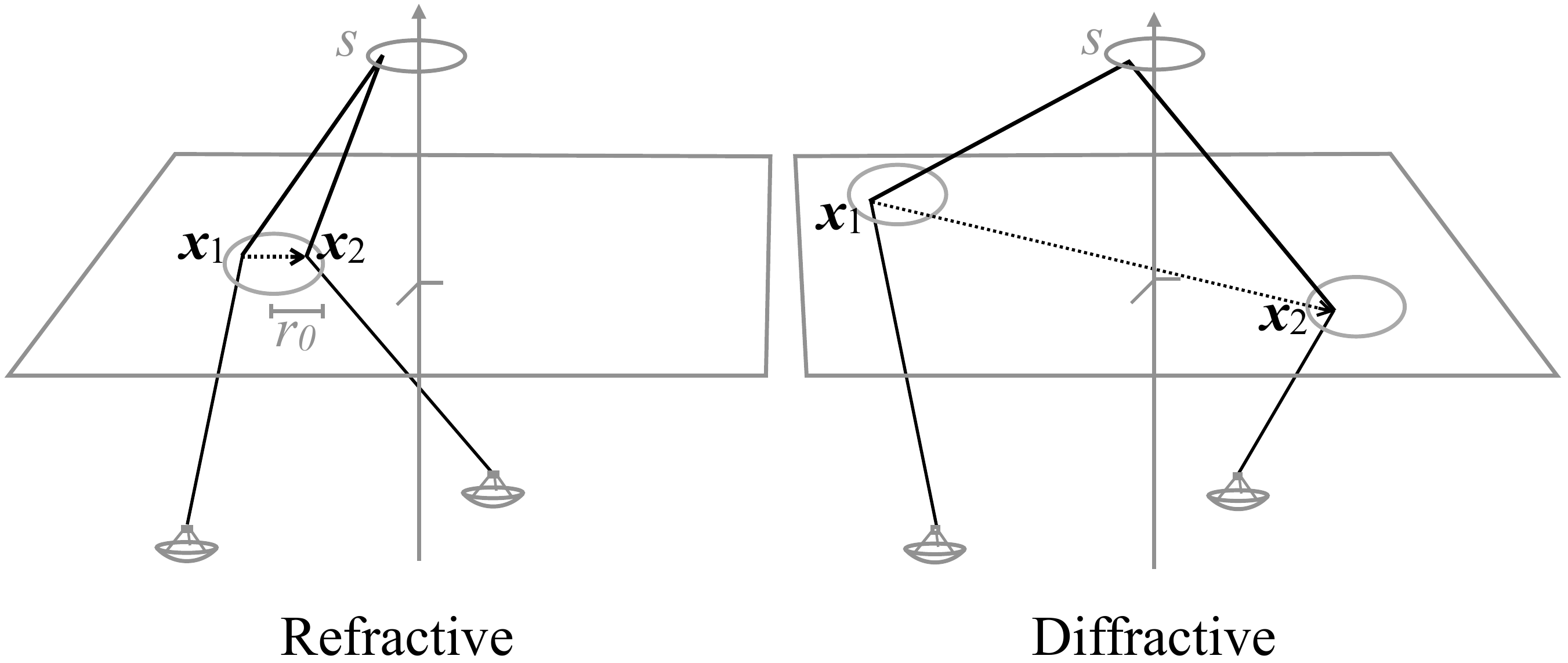}
\caption
{ 
Refractive and diffractive contributions to the snapshot visibility (Eq.~\ref{eq::Vsnapshot}), as discussed in \S\ref{sec::A_Visibility}.
}
\label{fig_refractive_diffractive_geometry}
\end{figure}

The snapshot visibility, corresponding to Eq.~\ref{eq::Vsnapshot}, represents the cumulative effect of three contributions: the ensemble-average visibility, diffractive noise, and refractive noise. As shown by \citetalias{GoodmanNarayan89} and illustrated in Figure~\ref{fig_refractive_diffractive_geometry}, the diffractive contribution to the integral comes from regions with $\left| \textbf{x}_1 - \textbf{x}_2 \right| \gg r_0$ (see Appendix \ref{sec::Refractive_Noise}); the ensemble average and refractive noise arise from the region with $\left| \textbf{x}_1 - \textbf{x}_2 \right| \lsim r_0$. Indeed, this conclusion is apparent because an extended source limits the contribution from widely-separated pairs of points and thereby quenches the diffractive scintillation \citep{ISO}. Thus, a extended source with size that is significantly larger than the diffractive scale will immediately take snapshot visibilities into the regime of the average visibilities. When the angular size of the source exceeds the 
refractive 
scale (i.e., when the scattering becomes subdominant to source structure), the 
source will impose an even tighter restriction $\left| \textbf{x}_1 - \textbf{x}_2 \right| \lsim (1+M)^{-1}r_{\rm img} \leq r_0$ that reflects the shorter correlation length of the source electric field at the screen. We define $r_{\rm img}$ more precisely in \S\ref{sec::Manifestations_of_Refractive_Noise}.

A coarse examination of Eq.~\ref{eq::Vsnapshot} reveals most salient properties of the noise in average visibilities. For instance, we can readily understand why average visibilities are quite insensitive to integration in frequency or time -- it is a consequence of the restriction $\left| \textbf{x}_1 - \textbf{x}_2 \right| \lsim (1+M)^{-1} r_{\rm img} \leq r_0$. For example, any significant effect from a shift of the baseline center $\Delta \textbf{b}_0$ requires $(1+M)^{-1} r_{\rm F}^{-2} \left| \Delta \textbf{b}_0 \right| \sqrt{ b^2 + (1+M)^{-2} r_{\rm img}^2} \gsim 1$. Thus, short baselines must be displaced by ${\sim} (1+M)^2 r_{\rm F}^2/r_{\rm img}$, which is the appropriately magnified size of the scattered image. Long baselines must be displaced by an amount that is smaller by $\sim (1+M)^{-1}r_{\rm img}/b$. Because baselines with $b \gg r_0$ resolve the ensemble-average scattering disk and, therefore, see little flux, the former is the more typical circumstance. Thus, to influence accessible 
refractive metrics, baseline centers 
must be displaced by the 
refractive scale. An extended source increases the spatial correlation scale, reinforcing the notion of a 
smooth diffractive pattern in the observing plane, the convolution of the source intensity distribution with the point-source response, as discussed in \S\ref{sec::Snapshot_Visibility}. 

We can likewise consider the effects of averaging in frequency. To incorporate frequency dependence, there are two necessary modifications: $r_{\rm F} \propto \sqrt{\lambda}$ and $\phi(\textbf{x}) \propto \lambda$. The additional phase from a wavelength change $\Delta \lambda$ is then given by $\frac{\Delta \lambda}{\lambda}\left\{ \frac{1}{2} r_{\rm F}^{-2}  \left(x_1^2 - x_2^2 \right) + \frac{1}{2} r_{\rm F}^{-2} (1+M)^{-1}\textbf{b}\cdot\left( \textbf{x}_1 + \textbf{x}_2 \right)+ \left[ \phi(\textbf{x}_1) - \phi(\textbf{x}_2) \right]\right\}$. 
In the refractive regime, $\left| \textbf{x}_1 - \textbf{x}_2 \right| \lsim r_0$, so $\phi(\textbf{x}_1) - \phi(\textbf{x}_2) \lsim 1$. 
Hence, the phase of the first term is ${\sim} \frac{\Delta \lambda}{\lambda}$, the second is ${\sim} \frac{\Delta \lambda}{\lambda} \frac{b}{r_{\rm img}}$, and the third is ${\sim} \frac{\Delta \lambda}{\lambda}$. Thus, the fractional bandwidth of refractive noise is of order unity, except on long baselines where it becomes ${\sim} r_{\rm img}/b$. Note that an extended source will \emph{decrease} the fractional bandwidth of refractive noise on long baselines but will not change the fractional bandwidth of refractive noise on short baselines. 

However, while the baseline center $\textbf{b}_0$ must shift by the refractive scale to incur a significant change in the average visibility, the vector baseline $\textbf{b}$ only needs to change by a distance of ${\sim} r_{\rm img}$. Thus, an extended source will \emph{decrease} the coherence length of the refractive noise in the visibility domain. This property already suggests the presence of intense substructure in the scattered image, which we explore in \S\ref{sec::Scattered_Image}.

Because the current paper emphasizes refractive effects, we will simply set:
\begin{align}
\textbf{b}_0 \equiv \textbf{0}
\end{align}
and ignore frequency dependence for the remainder.

\subsection{The Ensemble-Average Visibility}
\label{sec::EA_Visibility}

To obtain the ensemble-average visibility, we must average over many realizations of the scattering screen, defined by the Gaussian random field $\phi(\textbf{x})$. To calculate this average, the characteristic function of a zero-mean Gaussian random variable provides a convenient identity: $\left \langle e^{i \left[ \phi(\textbf{x}_1) - \phi(\textbf{x}_2) \right]} \right \rangle_{\rm ea} = e^{-\frac{1}{2} \left \langle \left[ \phi(\textbf{x}_1) - \phi(\textbf{x}_2) \right]^2 \right \rangle_{\rm ea}} = e^{-\frac{1}{2} D_{\phi}(\textbf{x}_1 - \textbf{x}_2)}$. 
We can then change the integration variables to $\textbf{y} \equiv \textbf{x}_1 - \textbf{x}_2$ and $\textbf{x} \equiv (\textbf{x}_1 + \textbf{x}_2)/2$ to give
\begin{align}
V_{\rm ea}(\textbf{b}) &= \frac{1}{4\pi^2 r_{\rm F}^4 } \int d^2 \textbf{y}\, d^2 \textbf{x}\, e^{i r_{\rm F}^{-2} \left[ \textbf{y} + \frac{\textbf{b}}{\left( 1 + M \right)} \right]\cdot\textbf{x}} e^{-\frac{1}{2} D_{\phi}\left( \textbf{y} \right) } V_{\rm src}(-\left(1+M\right)\textbf{y}).
\end{align}
Integration over $\textbf{x}$ is then trivial, using the identity $\int d^2\textbf{x}\, e^{i \textbf{y} \cdot \textbf{x}} = (2\pi)^2 \delta^2(\textbf{y})$. Thus, because $D_{\phi}(\textbf{y}) = D_{\phi}(-\textbf{y})$,
\begin{align}
V_{\rm ea}(\textbf{b}) = e^{-\frac{1}{2} D_{\phi}\left( \frac{\textbf{b}}{1+M} \right)} V_{\rm src}(\textbf{b}).
\end{align}
Hence, we recover the well-known result that the ensemble-average visibility for a scattered source is the product of the point source visibility and the unscattered source visibility. A familiar corollary is that, in the ensemble-average regime, scattering convolves the unscattered source image with a smooth scattering kernel. 

Indeed, because the ensemble-average scattering is deterministic and has a positive kernel, $e^{-\frac{1}{2}D_{\phi}\left(\frac{\textbf{b}}{1+M}\right)}$, it is readily inverted when known \citep{Fish_2014}. Hence, in the ensemble-average regime, one can reconstruct the \emph{unscattered} image of the source by dividing measured visibilities by the scattering kernel. With this approach, the residual effects of scattering are simply to amplify the thermal noise on long baselines. 

Moreover, the convolution action of scattering leads to other convenient properties in this regime, as discussed by \citet{Fish_2014} and \citet{Johnson_2014}. For instance, because the kernel is real and positive, visibility phases are unaffected by scattering in the ensemble-average regime, even if the scattering kernel is not known. Also, quotients of visibilities on equal baselines will be unaffected, such as fractional polarization in the visibility domain (because the scattering is not significantly birefringent).

\section{The Scattered Image}
\label{sec::Scattered_Image}

\subsection{Definition of the Scattered Image}

We now consider the appearance of the scattered image. However, there are two immediate difficulties in precisely defining the scattered image. The first is that it may depend so sensitively on observing position that any aperture sufficient to resolve the image will necessarily span different image elements. The second is that the scattered image is not incoherent, and so is not trivially related to visibilities via the Van Cittert-Zernike theorem. 

The first of these concerns is problematic for snapshot visibilities. However, we have seen that average visibilities are insensitive to changes of observing position (i.e., baseline center $\textbf{b}_0$) that are less than the refractive scale. Thus, the average image is well-defined in this regard. For the second concern, because the coherence length on the scattering screen is $\sim r_0 \ll r_{\rm R}$, the scattered image can be considered effectively incoherent for all baselines of length $\left| \textbf{b} \right| \ll \lambda D/r_0 \sim r_{\rm R}$. 

Hence, we can define the average image according to the Van Cittert-Zernike theorem applied to average visibilities:
\begin{align}
\label{eq::Image_Definition}
I_{\rm avg}(\textbf{x}) &\propto \int d^2\textbf{b}\, V_{\rm avg}(\textbf{b}) e^{-2\pi i \left(\frac{\textbf{b}}{\lambda}\right) \cdot \left(\frac{\textbf{x}}{D}\right)} = \int d^2\textbf{b}\, V_{\rm avg}(\textbf{b}) e^{-\frac{i}{r_{\rm F}^2} \frac{\textbf{b}}{1+M} \cdot \textbf{x}}.
\end{align}
Note that we have calculated the image at a distance of the scattering screen, $D$, rather than at a distance of the source. This convention was also adopted by \citetalias{NarayanGoodman89}, and allows linear coordinates to be easily compared on the screen. 
Note also that we ignore the dependence of the snapshot visibility on ${\bf b}_0$, in accord with the discussion of \S\ref{sec::A_Visibility} (a shift in $\textbf{b}_0$ is equivalent to a shift of the scattered image). 
Although this image is rather different from that produced by a physically realizable interferometer or aperture, 
it can be filtered or averaged to reproduce their behavior flexibly and easily.

Observe that the integrand of Eq.~\ref{eq::Image_Definition} is conjugated when $\textbf{b} \rightarrow -\textbf{b}$, so $I_{\rm avg}(\textbf{x}) \in \mathbb{R}$, although the image is not necessarily positive.

\subsection{The Scattered Image of an Extended Source}
\label{sec::Scattered_Image_Extended}

We now derive an expression for the scattered image of an extended source in the average-image regime. Here, we require only that the source quenches the diffractive scintillation so that the snapshot- and average-image visibilities are equal (see \S\ref{sec::Visibilities}). 
The preceding definition for the scattered image is especially useful in this case because the snapshot visibility has a straightforward dependence on baseline. After substituting $V_{\rm ss}(\textbf{b})$ (given by Eq.~\ref{eq::Vsnapshot}) for $V_{\rm avg}(\textbf{b})$ in Eq.~\ref{eq::Image_Definition}, the integral over $\textbf{b}$ gives a delta function with argument proportional to $\textbf{x} - (\textbf{x}_1 + \textbf{x}_2)/2$. Thus, the only contribution to the snapshot image at a location $\textbf{x}$ is from pairs of points on the screen that are centered on $\textbf{x}$. As in \S\ref{sec::EA_Visibility}, we can change to variables given by the average and difference of $\textbf{x}_1$ and $\textbf{x}_2$. Integrating over the former leaves a single remaining integral over $\textbf{y} \equiv \textbf{x}_2 - \textbf{x}_1$:
\begin{align}
\label{eq::Image_ss}
I_{\rm avg}(\textbf{x}) &\propto \int d^2\textbf{y}\, 
V_{\rm src}\left( (1+M) \textbf{y} \right)
e^{i \left[ \phi\left(\textbf{x}-{\textstyle{\frac{1}{2}}}\textbf{y}\right) - \phi\left(\textbf{x}+{\textstyle{\frac{1}{2}}}\textbf{y}\right)  \right]}
e^{ -\frac{i}{r_{\rm F}^{2}} \textbf{y}\cdot \textbf{x} }\!.
\end{align}
This form is especially convenient for numerical simulations. In \S\ref{sec::Image_Shuffle}, we will derive an even simpler approximate representation, which eliminates the remaining integral.

Note that if the screen-phase term $\exp\left\{i \left[ \phi\left(\textbf{x}-{\textstyle{\frac{1}{2}}}\textbf{y}\right) - \phi\left(\textbf{x}+{\textstyle{\frac{1}{2}}}\textbf{y}\right) \right]\right\}$ depended only on $\textbf{y}$, then $I_{\rm avg}(\textbf{x})$ would be the Fourier transform of the product of a position-independent scattering kernel and the source visibility (compare Eq.~\ref{eq::Image_ss} with Eq.~\ref{eq::Image_Definition});  consequently, the scattered image would be the convolution of the point-source response with an image of the source. 
Perhaps unfortunately, the screen-phase term depends on $\textbf{x}$ so the scattering does not act as a convolution for average images.  However, after an ensemble-average over the screen phases, 
 $\left \langle \exp\left\{i \left[ \phi(\textbf{x}-{\textstyle{\frac{1}{2}}}\textbf{y}) - \phi(\textbf{x}+{\textstyle{\frac{1}{2}}}\textbf{y}) \right]\right\} \right \rangle_{\rm ea} = e^{-\frac{1}{2} D_\phi(\textbf{y})}$,
 so that we again recover the expected convolution action of scattering in the ensemble-average regime.

\subsubsection{Existence and Persistence of Refractive Substructure}
\label{sec::existence_of_refractive substructure}

The arguments of the preceding section raise the question: why should refractive substructure exist at all?
If the screen phase, $\phi(\textbf{x})$, decorrelates over the scale of $r_0$ at the screen,
why does each small region of scale $r_0$ not produce an independent variation from the average image,
so that any larger-scale structures are completely random in character?  
As we will now demonstrate, although the phase decoheres over the scale $r_0$, the phase \emph{gradient} remains correlated over much larger scales.
This long-range correlation accounts for the existence and persistence of refractive substructure.

Specifically, note that the autocovariance of the directional derivative of phase can be related to the second derivative of the phase autocovariance:
\begin{align}
\label{eq::gradient_acf}
\left \langle \left[ \partial_x \phi\left(\textbf{x}_0 \right) \right] \left[ \partial_x \phi\left(\textbf{x}_0 + \textbf{x} \right) \right]\right \rangle = -\partial^2_{x} \left \langle \phi\left(\textbf{x}_0\right) \phi\left(\textbf{x}_0 + \textbf{x} \right) \right \rangle = \partial^2_{x} D_{\phi}(\textbf{x}).
\end{align}
Thus, for a power-law index $\alpha$, the phase-slope autocovariance only falls as $\left| \textbf{x} \right|^{-(2-\alpha)}$ (e.g., as $\left| \textbf{x} \right|^{-1/3}$ for a Kolmogorov spectrum). More precisely, note that the phase gradient causes a shift of the image by ${\sim} r_{\rm F}^2 \nabla \phi(\textbf{x})$ (we will show this explicitly in \S\ref{sec::Image_Shuffle}). Thus, the autocovariance of the fractional image shift, relative to the refractive scale, of points offset by $\textbf{x}$ is ${\sim}(\left| \textbf{x} \right|/r_0)^{-(2-\alpha)}$. A large inner scale extends this correlation length, while an outer scale introduces a sharp cutoff for the correlation. The broad correlation of phase gradient leads to the coherence of substructure on scales much larger than $r_0$ and allows it to persist over relatively long temporal averages which, with the assumptions of a ``frozen'' phase screen, are equivalent to partial spatial averages.

\subsubsection{The Role of Source Structure}
\label{sec::role_of_source}

We now address another fundamental question: does an extended source smooth the scattered image? To answer this, note that the effect of an extended source in Eq.~\ref{eq::Image_ss} is to restrict the range of $\textbf{y}$ to small values. For a given location $\textbf{x}$ on the image, this restriction decreases the variations of $\exp\left\{i \left[ \phi\left(\textbf{x}-{\textstyle{\frac{1}{2}}}\textbf{y}\right) - \phi\left(\textbf{x}+{\textstyle{\frac{1}{2}}}\textbf{y}\right) \right]\right\}$, which are the source of the refractive substructure (and the scatter broadening). 
This phase difference quickly approaches 0 when the angular size of the source exceeds the ensemble-average angular size of a scattered point source, thereby quenching the refractive substructure. 
However, because the image coordinate $\textbf{x}$ determines the center of the separation of contributing pairs of screen points in Eq.~\ref{eq::Image_ss} that produce the refractive substructure, the only role of source structure is to diminish the effects of this phase difference at each $\textbf{x}$ independently. 
Source structure will reduce the screen-induced phase fluctuations but will not spatially smooth them.

\begin{figure}[t]
\centering
\includegraphics*[width=\textwidth]{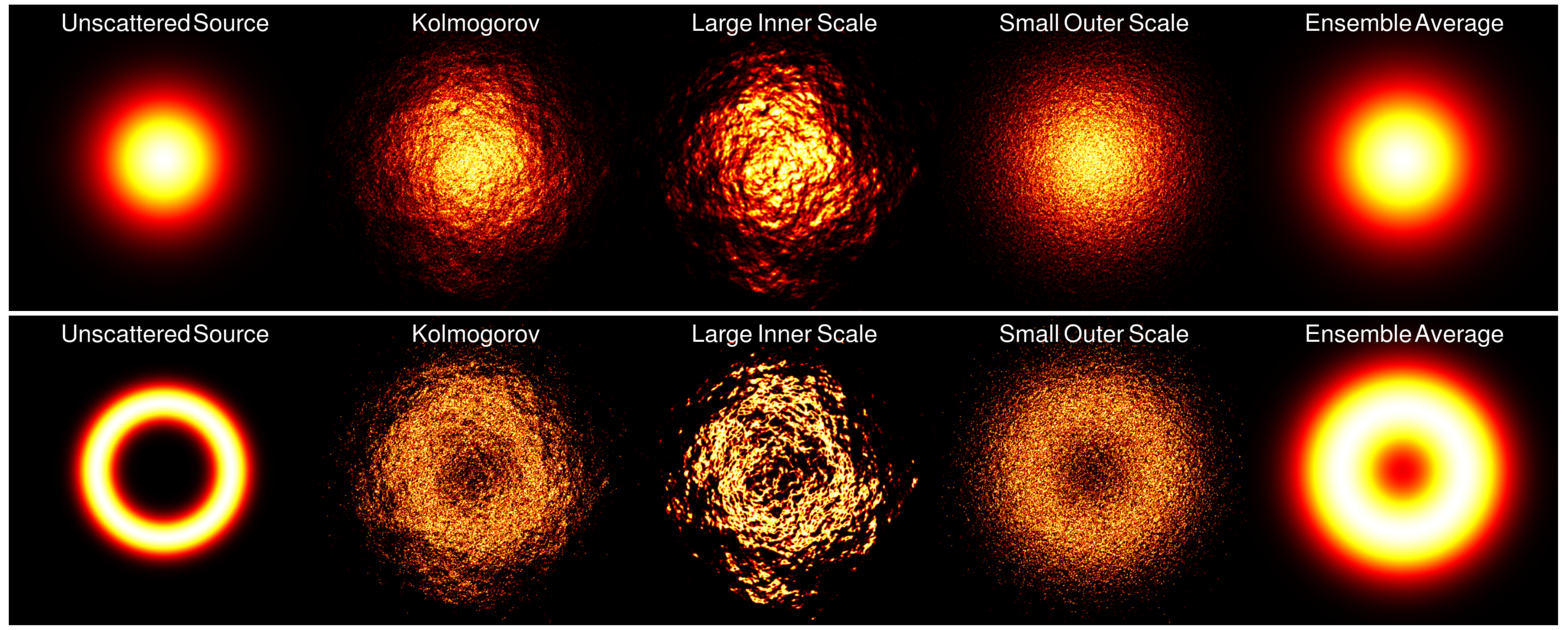}
\caption
{ 
Simulated images showing the effects of scattering for two sources: a circular Gaussian distribution ({\it top}) and a ring ({\it bottom}). The leftmost panel in each shows the unscattered source; the rightmost shows the ensemble-average scattered source. The three central panels show three different average images, each with a Kolmogorov index in the inertial range. The second and third have a large inner scale ($r_{\rm in} = 200 r_0 \approx 0.16 r_{\rm R}$) and a small outer scale ($r_{\rm out} = 500 r_0 \approx 0.41 r_{\rm R}$), respectively. Each image has $r_{\rm F} = 35 r_0$ and $M=1$ and was calculated using a scattering screen with $2^{14} \times 2^{14}$ random phases, with the same random seed for both source models. We used $r_{\rm in}=5r_0$ in each case (except the large inner scale examples) so that the description and numerical reduction of described in \S\ref{sec::Image_Shuffle} was appropriate. The color scale is linear and extends to the brightest pixel in each image; the angular size of each 
image is identical. Observe that the greater 
extent of the 
Gaussian than the ring results in a shallower 
signature of substructure in the average images but does not blur the substructure, in accord with the discussion of \S\ref{sec::role_of_source}. See Appendix~\ref{sec::Simulations} for additional details about the simulations.
}
\label{fig_image_examples}
\end{figure}

\subsubsection{Approximating the Scattered Image}
\label{sec::Image_Shuffle}

A limiting case provides considerable insight and a powerful tool for simplifying numerical work. Specifically, suppose that the screen phase fluctuations are smooth over the range of coherence for the source field at the screen: $V_{\rm src}\left( (1+M) r_{\rm in} \right) \ll 1$. Then, over the relevant range of integration in Eq.~\ref{eq::Image_ss},
  $\phi(\textbf{x}+\textstyle{\frac{1}{2}}\textbf{y}) - \phi(\textbf{x}-\textstyle{\frac{1}{2}}\textbf{y}) \approx \textbf{y} \cdot \nabla \phi(\textbf{x})$. With this approximation, Eq.~\ref{eq::Image_ss} becomes
\begin{align}
\label{eq::Image_Approx}
\nonumber I_{\rm ss}(\textbf{x}) &\approx \int d^2\textbf{y}\, 
V_{\rm src}\left( (1+M) \textbf{y} \right)
e^{-i \textbf{y} \cdot \nabla \phi(\textbf{x}) }
e^{-\frac{i}{r_{\rm F}^{2}} \textbf{y}\cdot \textbf{x} }\\
&=I_{\rm src}\left( \textbf{x} + r_{\rm F}^2 \nabla \phi(\textbf{x}) \right).
\end{align}
In this case, refraction ``shuffles'' the image, moving each image element to a place nearby: the scattered image brightness at a location $\textbf{x}$ is given by the brightness of the \emph{unscattered} source at $\textbf{x} + r_{\rm F}^2 \nabla \phi(\textbf{x})$, in accordance with the brightness theorem of geometrical optics \citep{Born_Wolf}. Since $|\nabla \phi| \sim 1/r_0$, the shuffling occurs over a region spanning the refractive scale, centered on $\textbf{x}$. Again, we emphasize that $\textbf{x}$ is a transverse coordinate at the scattering screen (not the source) with a corresponding angular coordinate $\textbf{x}/D$. 

This approximation vividly illustrates the breakdown of the convolution action of scattering in the average image, as discussed above. Conversely, in the ensemble-average limit, the averaged ``shuffling'' behavior leads to a smooth position-independent blurring around each image coordinate to reproduce the familiar convolution of point-source response and image, now in the ensemble-average regime.

Because the approximation of Eq.~\ref{eq::Image_Approx} eliminates the remaining position-dependent integral, it enables rapid estimation of the effects of scattering on any image after generating an appropriate random phase screen. Figure \ref{fig_image_examples} shows some examples of scattered images that utilize this approximation.

\begin{figure}[t]
\centering
\includegraphics*[width=\textwidth]{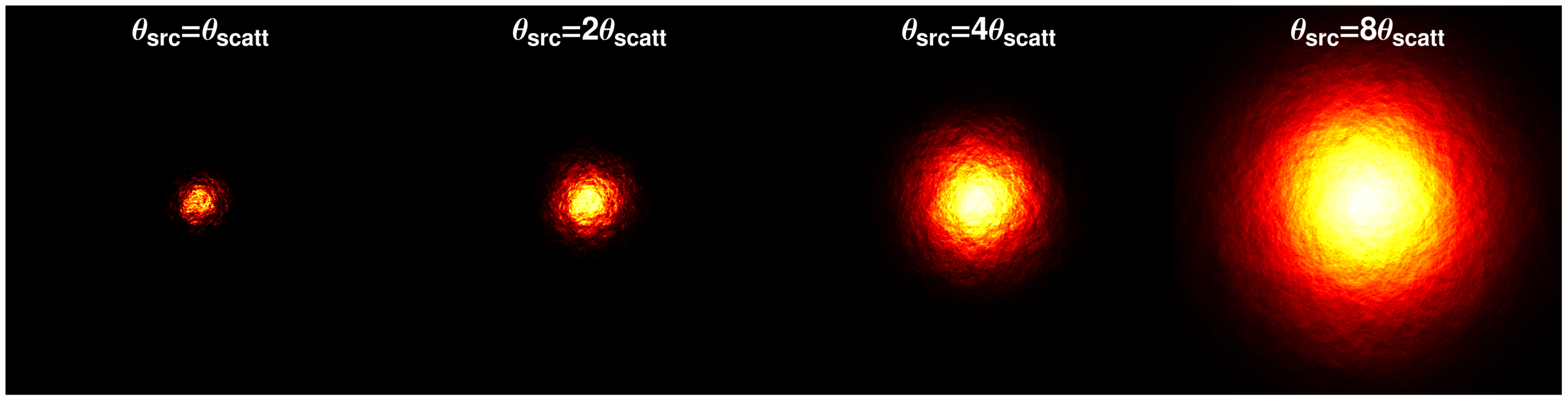}
\caption{
Simulated images showing the effects of source size on a scattered image of a circular Gaussian source, as in Figure~\ref{fig_image_examples}. Four cases are shown: the leftmost has a source that has the same angular size as the ensemble-average scattered image of a point source, the next three are for identical scattering but an intrinsic source that is 2, 4, and 8 times larger. In each case, the scattering is Kolmogorov with a small inner scale and infinite outer scale, a Fresnel scale $r_{\rm F} = 17.9 r_0$, a magnification $M=1$, and a scattering screen with $2^{14} \times 2^{14}$ random phases. The color scale is linear and extends to the brightest pixel in each image; the angular size of each image is identical. Even when the scattering is highly subdominant to the intrinsic structure, the effects of scattering-induced substructure are still readily apparent. 
}
\label{fig_image_sourcesize_examples}
\end{figure}

\subsection{The Imprint of Scattering Characteristics on the Scattered Image}
\label{sec::Image_Properties}

The discussion of \S\ref{sec::Scattered_Image_Extended} allows us to assess how the scattered image is modified by properties of the scattering (see Figure \ref{fig_image_examples} for examples). For instance, in \S\ref{sec::existence_of_refractive substructure} we showed that the large-scale coherent structures in the scattered image are related to the long-range correlation of the phase gradient $\nabla \phi(\textbf{x})$ across the image, which is proportional to the second derivative of $D_{\phi}(\textbf{x})$. 

Thus, if there is a finite outer scale, then the phase gradient \emph{must} decohere over displacements of ${\sim}r_{\rm out}$ to avoid excessive growth of the phase and divergence of the phase structure function. This decoherence quickly destroys the coherent features in the scattered image on scales larger than $r_{\rm out}$ and eliminates power on baselines too short to resolve the scale of the decoherence ($|\textbf{b}| \lsim \frac{\lambda D}{r_{\rm out}} \sim \frac{r_{\rm R}}{r_{\rm out}} r_0$). On the other hand, a large inner scale increases the correlation length of the phase gradient. This increase results in larger-scale coherent features in the scattered image, or increased ``patchiness.'' 

Similar considerations apply to the power-law index $\alpha$. Because larger values of $\alpha<2$ have an increased correlation length in the phase gradient (from Eq.~\ref{eq::gradient_acf}), they produce a higher level of refractive noise on short baselines but a steeper fall in the noise with increasing baseline.

\section{Manifestations of Refractive Noise}
\label{sec::Manifestations_of_Refractive_Noise}

We now discuss observable manifestations of the refractive noise, how each is quenched by an extended source, and how each reflects properties of the scattering. Our discussion relies on expressions for the variance of the snapshot visibility that are derived in Appendix \ref{sec::Refractive_Noise}. We outline how to obtain exact expressions numerically (using results in Appendix \ref{sec::RefractiveNoise}) but primarily focus on developing intuitive understanding, approximations, and scaling relationships. 

In \S\ref{sec::Flux_Modulation}, we discuss the most familiar refractive effect, modulation of the total image flux. Next, in \S\ref{sec::ImageWander}, we derive the refractive contributions to image wander and distortion. Finally, in \S\ref{sec::Refractive_Substructure_Image}, we derive expressions for the visibility noise on long baselines, reflecting refractive substructure in the scattered image. We summarize these results in \S\ref{sec::Noise_Summary} and in Table~\ref{tab::Refractive_Noise}.

\subsection{Refractive Flux Modulation}
\label{sec::Flux_Modulation}

The most familiar refractive manifestation, modulation of the total flux, has been studied in depth by many authors \citep[e.g.,][]{Shishov_1974,RCB_1984,Goodman_Narayan_1985}. Nevertheless, because the total flux is simply the visibility measured by a zero-baseline interferometer, we can use our formulation to estimate the flux modulation, providing a valuable point-of-contact with previous work. 

To derive an explicit estimation the flux modulation, we will approximate the ensemble-average image by a Gaussian with a characteristic scale $r_{\rm img}$ in the visibility domain: $V_{\rm ea}\left( \textbf{b} \right) \equiv e^{-\frac{1}{2}\left(\left|\textbf{b}\right|/r_{\rm img}\right)^2}$. The reciprocal of $r_{\rm img}$ is then proportional to the full width at half maximum (FWHM) of the scattered image, $\theta_{\rm img}$: $\theta_{\rm img} \sim \frac{\sqrt{2\ln 2}}{\pi} \frac{\lambda}{r_{\rm img}} \approx 0.37 \frac{\lambda}{r_{\rm img}}$. In turn, $\theta_{\rm img}$ is approximately given by the quadrature sum of the FWHM of the unscattered source, $\theta_{\rm src}$, and the FWHM of a point source, $\theta_{\rm scatt}$. 

With these substitutions, is it straightforward to evaluate Eq.~\ref{2D_general} on the zero-baseline. The rms flux variations are given by $\sqrt{\left\langle (\Delta  I)^2 \right \rangle} \propto r_{\rm F}^{\alpha-2} r_0^{-\alpha/2} r_{\rm img}^{2-\alpha/2} = (r_0/r_{\rm F})^{2-\alpha} \left(\theta_{\rm scatt}/\theta_{\rm img}\right)^{2-\alpha/2}$, in agreement with previous derivations \citep[e.g.,][]{Narayan_1992}. 

From Eq.~\ref{2D_general} and Eq.~\ref{eq::Q_isotropic}, it is evident that an inner scale only weakly affects the flux modulation until $r_{\rm in} \gsim r_{\rm R} r_0/r_{\rm img}$; the inner scale must be comparable to the transverse extent of the ensemble-average image to have a significant effect. However, the imprint of a small outer scale is much more severe -- as the outer scale becomes much less than $r_{\rm R}$, the rms flux modulation will be quenched as $\left(r_{\rm out}/r_{\rm R}\right)^{-(1+\alpha/2)}$ (see Eq.~\ref{eq::Q_general}).

\subsection{Refractive Image Wander and Distortion}
\label{sec::ImageWander}

Observations that are sensitive to the absolute image position will see refractive image wander, and observations on short baselines that begin to resolve the scattered image will see large-scale image distortion. In general, these effects will be correlated with each other and with the flux modulation \citep{Blandford_Narayan_1985,RNB_1986}. We will now show that these effects are quite generally related to refractive noise on short interferometric baselines, and we will then use the results of Appendix~\ref{sec::RefractiveNoise} to estimate their magnitude. 

Working in dimensionless baseline $\textbf{u} \equiv \textbf{b}/\lambda$ for the visibility $V(\textbf{u})$ and angular coordinates for the source image $I(\boldsymbol{\theta})$, the Van Cittert-Zernicke Theorem takes the form
\begin{align}
V(\textbf{u}) &= \int d^2\boldsymbol{\theta}\, I(\boldsymbol{\theta}) e^{2\pi i \textbf{u} \cdot \boldsymbol{\theta}}. 
\end{align}
For a short baseline -- i.e., a baseline that does not resolve the ensemble-average image, so that $|\textbf{u}| \ll 1/\theta_{\rm img}$ -- we can approximate the visibility by expanding the exponential in $2\pi i \textbf{u} \cdot \boldsymbol{\theta}$:
\begin{align}
\label{eq::Centroid}
V(\textbf{u}) &\approx  \llangle I(\boldsymbol{\theta})  \rrangle + 2\pi i \textbf{u} \cdot  \llangle \boldsymbol{\theta}\,I(\boldsymbol{\theta})  \rrangle - 2\pi^2  \llangle[\Big]  \left(\textbf{u} \cdot \boldsymbol{\theta}\right)^2\,I(\boldsymbol{\theta}) \rrangle[\Big].
\end{align}
Here, $\llangle f(\boldsymbol{\theta}) \rrangle \equiv \int d^2\boldsymbol{\theta}\, f(\boldsymbol{\theta})$ denotes an angular average. Likewise, the visibility noise $\Delta V(\textbf{u})$ is related to the image noise $\Delta I(\boldsymbol{\theta})$ as
\begin{align}
\label{eq::Noise_Short_Baseline}
\Delta V(\textbf{u}) &\approx \underbrace{ \vphantom{ \llangle[\Big]\rrangle[\Big] } \llangle \Delta I(\boldsymbol{\theta}) \rrangle}_{\substack{\text{Flux}\\\text{Modulation}}} 
+ \underbrace{ \vphantom{ \llangle[\Big]\rrangle[\Big] } 2\pi i \textbf{u} \cdot \llangle \boldsymbol{\theta}\,\Delta I(\boldsymbol{\theta}) \rrangle}_{\text{Image Wander}} 
- \underbrace{2\pi^2 \llangle[\Big]  \left(\textbf{u} \cdot \boldsymbol{\theta}\right)^2\,\Delta I(\boldsymbol{\theta}) \rrangle[\Big]}_{\text{Image Distortion}}. 
\end{align}
Thus, on short baselines, image wander will affect the visibility \emph{phase} while image distortion will affect the visibility \emph{amplitude}. 
Noise in $|V(\textbf{u})|$ will be quadratic in baseline:
\begin{align}
\left \langle \left| \Delta V(\textbf{u}) \right|^2 \right \rangle &\approx \llangle \Delta I(\boldsymbol{\theta})\rrangle^2 \left[ 1 + 4\pi^2 u^2 \left\{ \left[\hat{\textbf{u}} \cdot \frac{\llangle \boldsymbol{\theta}\,\Delta I(\boldsymbol{\theta})\rrangle}{\llangle \Delta I(\boldsymbol{\theta}) \rrangle}\right]^2 
- \frac{\llangle[\Big]  \left(\hat{\textbf{u}} \cdot \boldsymbol{\theta}\right)^2\,\Delta I(\boldsymbol{\theta}) \rrangle[\Big]}{ \llangle \Delta I(\boldsymbol{\theta}) \rrangle} \right\} \right].
\end{align}
Note that the quadratic coefficient comes from the combined effects of image wander and image distortion, which act in opposition. The two effects could be decoupled by evaluating the second moment of the snapshot visibility, $\left \langle V_{\rm ss}^2 \right \rangle$, in addition to the second moment of the snapshot visibility modulus (see Appendix \ref{sec::Vss}), for instance. Figure~\ref{fig_noise_cartoon} illustrates these changing contributions to noise in the snapshot visibility as a function of baseline length.

\begin{figure}[t]
\centering
\includegraphics*[width=0.9\textwidth]{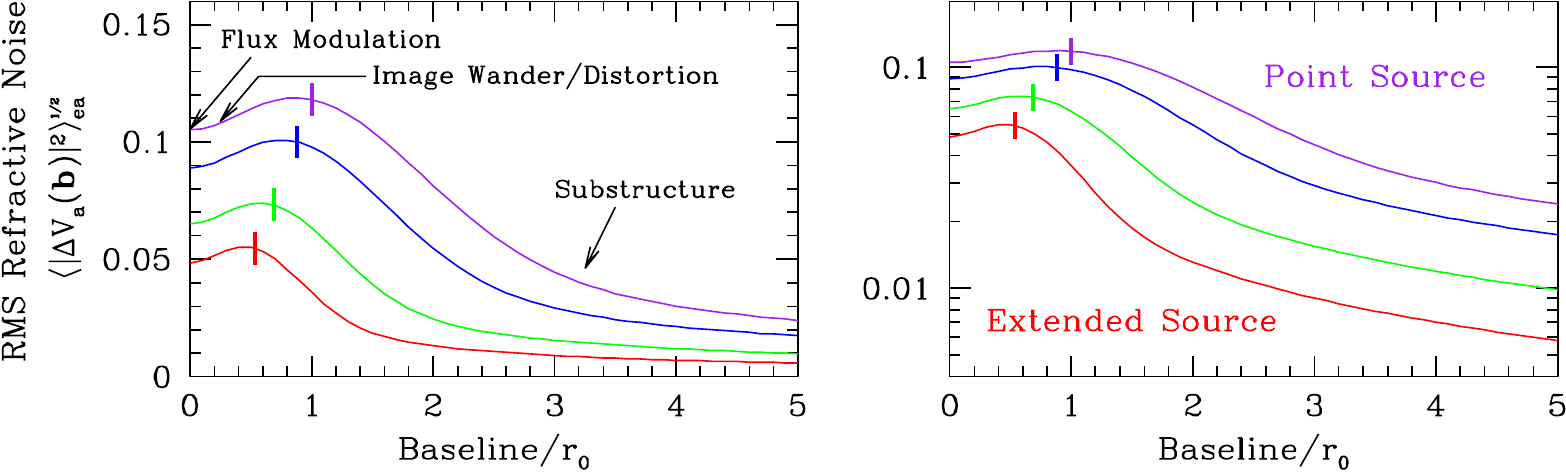}
\caption
{ 
Fractional rms refractive noise (Eq.~\ref{2D_general}) as a function of baseline for four values of linearly incremented source size, beginning with a point source. Left panel shows a linear scale, right shows logarithmic. The scattering parameters corresponding to these curves are $r_{\rm F} = 100 r_0$, $M=0$, and $\alpha=5/3$; the sources are circular Gaussians with unscattered sizes that are approximately 0, 0.5, 1, and 1.5 times the scattered size of a point source. On each curve, the heavy tick mark indicates the baseline on which the ensemble average visibility falls to $1/\sqrt{e}\approx 0.61$; for a point source, this baseline corresponds to $(1+M)r_0$. In each case, this location is near the peak of the refractive noise curve. The three noise regimes discussed in \S\ref{sec::Manifestations_of_Refractive_Noise} are indicated: 1) zero-baseline noise reflects refractive flux modulation; 2) short baselines see additional noise, increasing with the square of baseline until turning over at the noise peak, 
from the combined effects of image wander and distortion; 
3) long baselines resolve the ensemble-average image 
and are subject to noise from scattering-induced substructure within the average image. An increasing source size affects the noise differently in these regimes: flux modulation falls as $\theta_{\rm img}^{-(2-\alpha/2)}$, fractional image wander and distortion also as $\theta_{\rm img}^{-(2-\alpha/2)}$, and substructure as $\theta_{\rm img}^{-2}$. 
}
\label{fig_noise_cartoon}
\end{figure}

By expanding Eq.~\ref{2D_general} in baseline, we can now easily estimate the relative strength of the position wander and image distortion. Again adopting a circular Gaussian source with characteristic scale $r_{\rm img}$ in the visibility domain ($\theta_{\rm img} \sim 0.37 \lambda/r_{\rm img}$), we find that the quadratic coefficient has a relative amplitude of $\approx \frac{\alpha^2-2}{(2-\alpha)} (\lambda/r_{\rm img})^2$; this coefficient is actually quite a steep function of $\alpha$, ranging from $2.3$ at $\alpha=5/3$ to $16.1$ at $\alpha=1.9$. The aggregate rms position wander and distortion (in radians), $\sqrt{\left \langle \Delta \theta^2 \right \rangle}$, along one axis is then roughly
\begin{align}
\sqrt{\left \langle \Delta \theta^2 \right \rangle} = \frac{\lambda}{r_{\rm img}}\sqrt{\frac{\alpha^2-2}{4\pi^2 (2-\alpha) } \left \langle \Delta I^2 \right \rangle} \propto \left( \frac{r_0}{r_{\rm F}} \right)^{2-\alpha} \left( \frac{\theta_{\rm scatt}}{\theta_{\rm img}} \right)^{1-\alpha/2} \left( \frac{\lambda}{r_0} \right).
\end{align}
Note that, for a point source, this scaling matches the result of \citep{CPL_1986} (our $\alpha$ is their $\alpha-2$) and \citet{RNB_1986} (our $\alpha$ is their $\beta-2$). Interestingly, while most refractive effects tend to weaken with increasing wavelength, absolute image wander and distortion become stronger (see Table~\ref{tab::Refractive_Noise}). 

It is perhaps more natural to express $\sqrt{\left \langle \Delta \theta^2 \right \rangle}$ as a fraction of the angular size of the ensemble-average image, $\frac{r_{\rm R}}{D} \frac{\theta_{\rm img}}{\theta_{\rm scatt}} \sim \frac{\lambda}{r_0} \frac{\theta_{\rm img}}{\theta_{\rm scatt}}$. Then, the rms fractional wander is $\propto \left( \frac{r_0}{r_{\rm F}} \right)^{2-\alpha} \left( \frac{\theta_{\rm scatt}}{\theta_{\rm img}} \right)^{2-\alpha/2}$ -- the same scaling as the flux modulation. But most of the decrease in fractional wander for an extended source comes from the increasing size of the ensemble-average image. For instance, for a Kolmogorov spectrum, the rms wander (in radians) is only suppressed by $\left( \frac{\theta_{\rm scatt}}{\theta_{\rm img}} \right)^{1/6}$ relative to the wander of a point source. Hence, refractive image wander may remain significant even when the refractive flux modulation is heavily quenched.

\subsection{Refractive Substructure in the Scattered Image}
\label{sec::Refractive_Substructure_Image}

Finally, observations that can resolve the scattered image will see substructure introduced by scattering. This substructure contributes noise to long-baseline visibilities that is persistent and wideband, in accord with the conditions derived in \S\ref{sec::A_Visibility}. 
On a fixed baseline that is long enough to resolve the ensemble-average image (i.e., $\left| \textbf{b} \right| \gg r_{\rm img}$), Eq.~\ref{2D_general} shows that 
\begin{align}
\left \langle |\Delta V_{\rm a}(\textbf{b})|^2 \right \rangle_{\rm ea} &\propto \tilde{D}_{\phi}\left( \frac{\textbf{b}}{1+M} \right) \int d^2 \textbf{y}_{2}\, \left( \textbf{y}_2 \cdot \textbf{b}\right)^2 \left| V_{\rm ea}\left( (1+M)\textbf{y}_2 \right) \right|^2\\
\nonumber &\propto \tilde{D}_{\phi}\left( \frac{\textbf{b}}{1+M} \right)\int d^2 \boldsymbol{\theta}\,  \left| \left(\textbf{b} \cdot \nabla \right) I_{\rm ea}\left( \boldsymbol{\theta} \right) \right|^2.
\end{align}
This form highlights the major features of substructure on long baselines: the rms noise falls with baseline as $\left|\textbf{b}\right|^{-\alpha/2}$, the fractional suppression of the noise due to an extended source is independent of baseline length but may depend on the baseline orientation, and the noise is proportional to the root mean squared gradient of the unscattered source brightness. The refractive noise is then inversely proportional to the squared size of the ensemble-average image. For the specific case of source and scattering that are circularly-symmetric Gaussians, the rms refractive noise for long-baseline visibilities is $\sqrt{\left \langle \left| \Delta V_{\rm ss}(\textbf{b}) \right|^2 \right\rangle} \propto \left( \frac{r_0}{r_{\rm F}} \right)^{2-\alpha} \left( \frac{\left| \textbf{b} \right|}{r_0} \right)^{-\alpha/2} \left( \frac{\theta_{\rm scatt}}{\theta_{\rm img}} \right)^2$. 

From Eq.~\ref{2D_general}, we can also quickly understand how the inner and outer scales will affect the long-baseline noise. For instance, a finite outer scale will have little effect on the refractive noise on baselines $ \left| \textbf{b} \right| \gsim (1+M) r_{\rm F}^2 r_{\rm out}^{-1} = \frac{\lambda D}{2\pi} r_{\rm out}^{-1} = (1+M) r_0 (r_{\rm R}/r_{\rm out})$. These baselines are long enough to resolve the decoherence length, $r_{\rm out}$, set by the outer scale. Likewise, a finite inner scale introduces a sharp cutoff in the noise for baselines longer than $(1+M)\frac{r_0}{r_{\rm in}} r_{\rm R}$. This cutoff arises because the scattered image consists of many magnified and distorted ``subimages'' of the unscattered source, with a characteristic scale $r_{\rm in}$.\footnote{We thank Ramesh Narayan for identifying this correspondence.} Thus, the visibility noise falls sharply on baselines long enough to resolve the individual subimages.

A final important consideration is anisotropic scattering. Consider an anisotropic Gaussian scattering disk with an axial ratio $A\geq1$. By substituting Eq.~\ref{Dtile_Anisotropic} into Eq.~\ref{2D_general}, we find that along the major axis of the scattering disk, the refractive noise is enhanced by a factor of $A$ relative to that for a circular scattering disk with diameter equal to that of the major axis; likewise, along the minor axis, the noise is reduced by a factor $1/A$ relative to circular scattering with diameter equal to that of the minor axis. \citep{Johnson_Thesis}. The asymptotic ratio of the rms refractive noise along the major axis to the rms noise along the minor axis is then $A^{\alpha/2}$.

\subsection{Summary of Refractive Noise}
\label{sec::Noise_Summary}

{
\begin{deluxetable}{l||c|cc}
\tablewidth{\textwidth}
\tablecaption{Manifestations of Refractive Noise.}
\tablehead{
\colhead{}  & \colhead{}  & \multicolumn{2}{c}{Wavelength Dependence (Point Source)}  \\
\colhead{ \textbf{Metric} } & \colhead{ Approximate RMS } & \colhead{ Fixed Baseline } & \colhead{ Fixed Resolution }   }
\startdata
Fractional Flux Modulation         & $\left(\frac{r_0}{r_{\rm F}}\right)^{2-\alpha} \left(\frac{\theta_{\rm scatt}}{\theta_{\rm img}}\right)^{2-\alpha/2}$ & \multicolumn{2}{c}{$\lambda^{-\left( \frac{4}{\alpha} - \frac{\alpha}{2} - 1\right)}$} \\
Image Wander \& Distortion         & $\left( \frac{\lambda}{r_0} \right)\left( \frac{r_0}{r_{\rm F}} \right)^{2-\alpha} \left( \frac{\theta_{\rm scatt}}{\theta_{\rm img}} \right)^{1-\alpha/2} $ & \multicolumn{2}{c}{$\lambda^{2+\frac{\alpha}{2} - \frac{2}{\alpha}}$} \\
Refractive Substructure      & $ \left( \frac{\left|\textbf{b}\right|}{(1+M)r_0} \right)^{-\alpha/2} \left( \frac{r_0}{r_{\rm F}} \right)^{2-\alpha}\left( \frac{\theta_{\rm scatt}}{\theta_{\rm img}} \right)^{2}$ & $\lambda^{-\left( \frac{4}{\alpha} - \frac{\alpha}{2}\right)}$ & $\lambda^{-\frac{4}{\alpha}}$ 
\enddata
\label{tab::Refractive_Noise}
\end{deluxetable}
}

To summarize, visibilities on different baselines are sensitive to various manifestations of refractive noise. A zero-baseline interferometer sees fluctuations that correspond to flux modulation of the image. Short baselines, which do not resolve the scattered image, see an additional contribution: the aggregate of image wander and large-scale image distortion. The rms noise peaks near baselines of length $r_{\rm img}$ where $V_{\rm ea}\left(r_{\rm img}\right) \equiv 1/\sqrt{e}$ and then transitions to a power-law regime on longer baselines, with the visibility noise rms falling with baseline as $\left|\Delta V\right| \propto \left|\textbf{b}\right|^{-\alpha/2}$, reflecting compact substructure in the scattered image. The length $r_{\rm img}$ also determines the correlation length of the visibility noise in the observing plane: vector baselines that differ by more than a distance of $r_{\rm img}$ will see different realizations of the noise. See Table~\ref{tab::Refractive_Noise} for the behavior of the 
noise in each of these regimes.

Because these manifestations of refractive noise are affected differently by source structure, they provide a way of decoupling the source structure from the scattering without requiring multi-frequency measurements. For instance, \citet{Gwinn_2014} used the level of substructure for \sgra\ at $\lambda=1.3~{\rm cm}$ on long baselines to constrain the pair of parameters $\{ \alpha, \theta_{\rm src} \}$ and found good agreement with values that have been inferred by extrapolating scaling laws from much longer wavelengths. Moreover, some refractive effects may be much easier to detect than others. For instance, refractive image wander is minimally affected by an extended source whereas flux modulation falls quite steeply with increasing source size. Furthermore, the effects of an inner and outer scale depend strongly on baseline. Thus, a comprehensive study of refractive effects can provide a robust understanding of the source and scattering characteristics.

\section{Observational Implications}
\label{sec::Observational_Implications}

Our primary result is that small-scale refractive substructure persists in the presence of an extended source, even on baselines that would completely resolve the \emph{unscattered} source. Consequently, very-long baseline interferometry (VLBI) might detect structures smaller than any for the unscattered source, even if scattering does not significantly blur the source. We now give example calculations and specific implications for two ongoing projects: RadioAstron and the Event Horizon Telescope.

\subsection{Implications for RadioAstron}

\afterpage{
\begin{figure}[t]
\centering
\includegraphics*[width=0.9\textwidth]{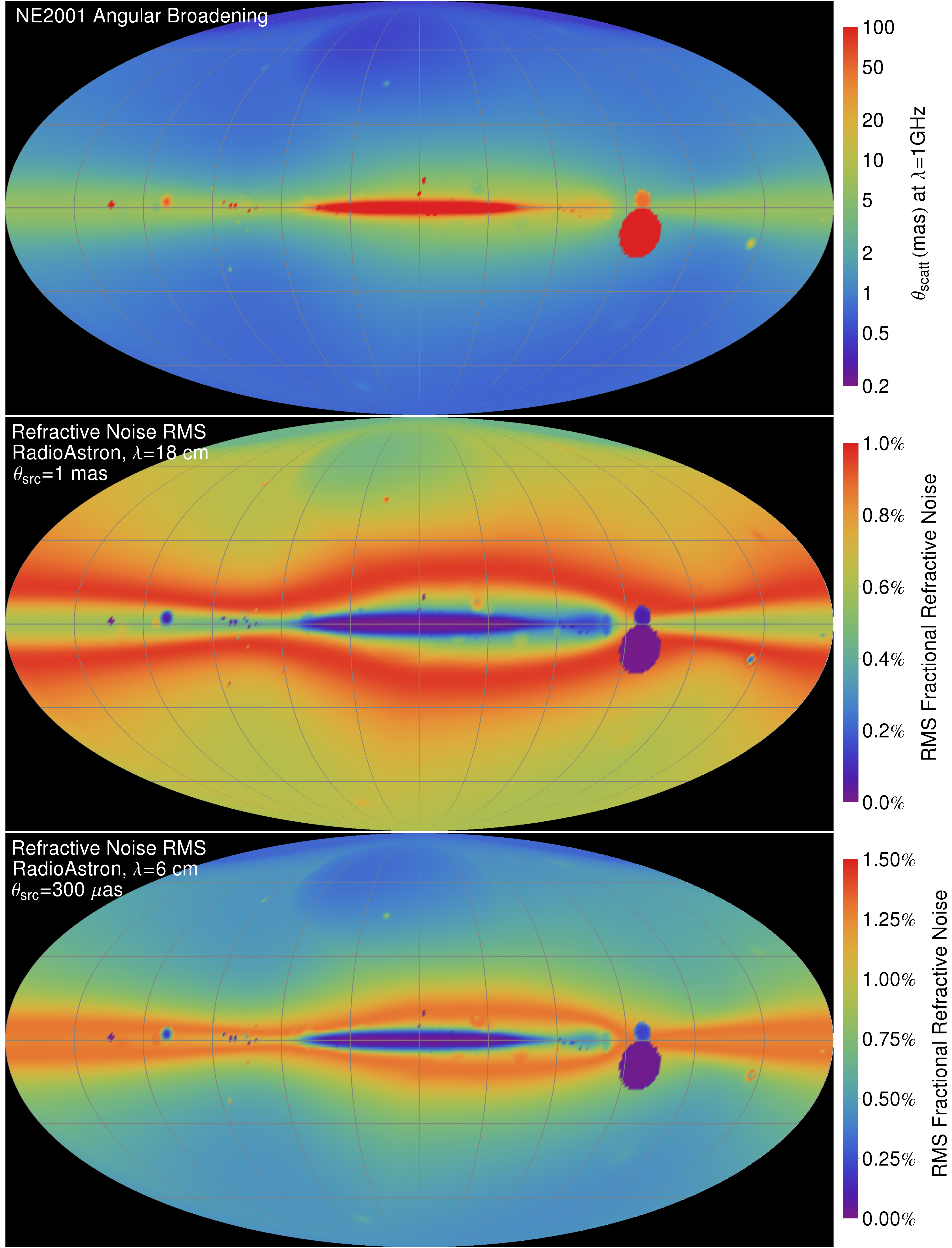}
\caption
{ 
Implications of refractive substructure for the RadioAstron mission; angular broadening ({\it top}) and refractive noise ({\it middle}, {\it bottom}) on a 100{,}000-km baseline; all images utilize a Mollweide projection. ({\it top}) Prediction of angular broadening at $\lambda = 1~{\rm GHz}$ for extragalactic objects according to the NE2001 model \citep{NE2001}; the median angular broadening is ${\sim}1~{\rm mas} \times \lambda_{\rm GHz}^{-2}$. ({\it middle}) Refractive noise rms, relative to the zero-baseline flux, at $\lambda=18~{\rm cm}$ on a 100{,}000-km baseline for a circular Gaussian source with FWHM of $\theta_{\rm src} = 1~{\rm mas}$. For comparison, the median angular broadening is $\theta_{\rm scatt} \approx 300\ \mu{\rm as}$. ({\it bottom}) 
Refractive noise rms, relative to the zero-baseline flux, at $\lambda=6~{\rm cm}$ on a 100{,}000-km baseline for a circular Gaussian source with FWHM of $\theta_{\rm src} = 300~\mu{\rm as}$. For comparison, the median angular broadening is $\theta_{\rm scatt} \approx 30\ \mu{\rm as}$. 
In each case, the rms noise varies as $\theta_{\rm img}^{-2} \approx \left(\theta_{\rm src}^{2} + \theta_{\rm scatt}^{2}\right)^{-1}$.
}
\label{fig_RadioAstron}
\end{figure}
\clearpage
}

RadioAstron is a 10-m radio dish in a highly eccentric elliptical orbit around the Earth \citep{Kardashev_2013}. Launched in July 2011 and operating at wavelengths ranging from 1.2 cm to 92 cm with Earth-space baselines as long as $370{,}000~{\rm km}$, it provides unprecedented angular resolution at these frequencies.
A key science goal of RadioAstron is to investigate the brightness temperatures of the most compact active galactic nuclei (AGN) via direct imaging. Since brightness temperature $T_{\rm b} \propto \left(\lambda/\theta_{\rm src}\right)^{2}$, the limiting brightness temperature that a baseline can detect is proportional to its length squared but not to its frequency \citep{Kovalev_2005}.
However, the long-baseline coverage of RadioAstron is quite sparse because the single orbiting antenna effectively provides only a single Earth-space baseline. Measurements on this baseline are by far the most important contributor when inferring compact structures, so it is essential to quantify potential refractive substructure. We will now show that refractive noise can be mistakenly identified as a signature of compact intrinsic structure, leading to significant over-estimates of intrinsic brightness temperature.  In contrast, scatter-broadening will always lead to an overestimation of the intrinsic size and, thus, an underestimation of the intrinsic brightness temperature. 

To derive a suitable point-of-reference, we consider a pure Kolmogorov structure function. 
We suppose that scattering material lies within the Milky Way, at a typical distance $D\approx 1\ {\rm kpc}$, and that the source is at cosmological distance so that $M=0$.
A typical long RadioAstron  baseline has length $b \sim 10^5\ {\rm km}$. 
We take a typical source angular size to be 
$\theta_{\rm src} \approx 300\ \mu{\rm as}$ at $\lambda=6\ {\rm cm}$,  
$\theta_{\rm src} \approx 1000\ \mu{\rm as}$ at $\lambda=18\ {\rm cm}$, and 
$\theta_{\rm src} \approx 5000\ \mu{\rm as}$ at $\lambda=92\ {\rm cm}$. These values are chosen to be somewhat larger than the most compact sources that RadioAstron can resolve at each frequency; the resultant $\theta_{\rm src} \propto \lambda$ scaling is also prototypical for self-absorbed jets \citep{Blandford_Konigl_1979} and our chosen sizes are typical of compact AGN \citep[e.g.,][]{Pushkarev_Kovalev_2015}. 
Figure~\ref{fig_RadioAstron} shows the estimated angular broadening using the NE2001 model \citep{NE2001};
we take as typical values $\theta_{\rm scatt}\approx  30~\mu{\rm as}$ at $\lambda=6~{\rm cm}$, $\theta_{\rm scatt} \approx 300~\mu{\rm as}$ at $\lambda=18~{\rm cm}$, and 
$\theta_{\rm scatt} \approx 8000~\mu{\rm as}$ at $\lambda=92~{\rm cm}$, which are approximately the median of the NE2001 angular broadening and, thus, represent typical lines of sight away from the Galactic plane. 
The size of the observed, scatter-broadened image is $\theta_{\rm img}\approx \sqrt{\theta_{\rm src}^2 + \theta_{\rm scatt}^2}$.

Under these assumptions, the fractional rms visibility from refractive noise, $\sigma_{\rm ref}(\textbf{b}) \equiv \sqrt{\left \langle \left| \Delta V_{\rm ss}(\textbf{b}) \right|^2 \right\rangle}$, in the RadioAstron observing bands is approximately:\footnote{At $\lambda=1.2~{\rm cm}$, the median angular broadening predicted by the NE2001 model is only ${\sim}1~\mu{\rm as}$, resulting in $r_0 \sim r_{\rm F}$ for a screen at $D=1~{\rm kpc}$. Thus, most observations with RadioAstron at $\lambda=1.2~{\rm cm}$ will be in a strong/weak scattering transition regime and Eq.~\ref{eq::RA_Estimate_1} is not directly applicable.}
\begin{align}
\label{eq::RA_Estimate_1}
\sigma_{\rm ref}(\textbf{b}) 
&= 0.0038 \times \left( \frac{\lambda}{6~{\rm cm}} \right) \left( \frac{\left| \textbf{b} \right|}{10^5~{\rm km}} \right)^{-5/6} \left( \frac{\theta_{\rm scatt}}{30~\mu{\rm as}} \right)^{5/6} \left( \frac{\theta_{\rm img}}{300~\mu{\rm as}}\right)^{-2} \left( \frac{D}{1~{\rm kpc}} \right)^{-1/6}\\
\nonumber &= 0.0071 \times \left( \frac{\lambda}{18~{\rm cm}} \right) \left( \frac{\left| \textbf{b} \right|}{10^5~{\rm km}} \right)^{-5/6} \left( \frac{\theta_{\rm scatt}}{300~\mu{\rm as}} \right)^{5/6} \left( \frac{\theta_{\rm img}}{1000~\mu{\rm as}}\right)^{-2} \left( \frac{D}{1~{\rm kpc}} \right)^{-1/6}\\
\nonumber &= 0.0055 \times \left( \frac{\lambda}{92~{\rm cm}} \right) \left( \frac{\left| \textbf{b} \right|}{10^5~{\rm km}} \right)^{-5/6} \left( \frac{\theta_{\rm scatt}}{8000~\mu{\rm as}} \right)^{5/6} \left( \frac{\theta_{\rm img}}{10000~\mu{\rm as}}\right)^{-2} \left( \frac{D}{1~{\rm kpc}} \right)^{-1/6}\!,
\end{align}
where we have applied the relationship between $r_0$ and $\theta_{\rm scatt}$ for $M=0$:
\begin{align}
\theta_{\rm scatt} \approx \frac{\sqrt{2\ln 2}}{\pi} \frac{\lambda}{r_0} \approx 0.37 \frac{\lambda}{r_0}.
\end{align}
A typical bright, compact source observed by RadioAstron might have a flux density in its compact component of $1-10$~Jy,
so we might expect substructure with strength $5-50$~mJy. Because baselines from RadioAstron to sensitive Earth antennas such as Arecibo have 4-$\sigma$ detection sensitivities $\lsim 10~{\rm mJy}$, substructure is quite plausibly detectable, even on baselines that would completely resolve the unscattered source.\footnote{For estimates of the RadioAstron SEFD in each observing band, see the RadioAstron User Handbook (\url{www.asc.rssi.ru/radioastron/documents/rauh/en/rauh.pdf}) or \citet{Kovalev_2014}.} 
Figure\ \ref{fig_RadioAstron} shows $\sigma_{\rm ref}(\textbf{b})$ as a function of position
on the sky, using the NE2001 model \citep{NE2001}.

It is evident that long baseline detections by RadioAstron must be interpreted with caution and compared with expected levels of refractive noise. Refractive noise would be constant over the refractive timescale of hours to weeks depending on the observing frequency and line of sight, and would be constant across the bandwidth of RadioAstron. It could be correlated among vector baselines that differ by up to a few times $10{,}000$ km. Over many observing epochs, the visibility amplitudes on a fixed baseline would be drawn from a Rayleigh distribution; they would have a mean value of $\frac{\sqrt{\pi}}{2} \sigma_{\rm ref}(\textbf{b})$ with 95\% of samples drawn from the range $[0.16, 1.92]\times \sigma_{\rm ref}(\textbf{b})$. To determine whether a measurement of $\theta_{\rm src}$ is a secure indication of compact intrinsic structure, one must use Eq.~\ref{eq::RA_Estimate_1} together with best estimates of $\theta_{\rm scatt}$ along the particular line of sight, to see whether the measured long-baseline 
visibilities can be reproduced from refractive noise with a larger $\theta_{\rm src}$ than is inferred in the absence of refractive noise. Long-baseline measurements that are consistently \emph{lower} than the expected refractive noise could be used to infer extended structure that quenches refractive noise.

\subsection{Implications for the Event Horizon Telescope}

\begin{figure}[t]
\centering
\includegraphics*[width=1.0\textwidth]{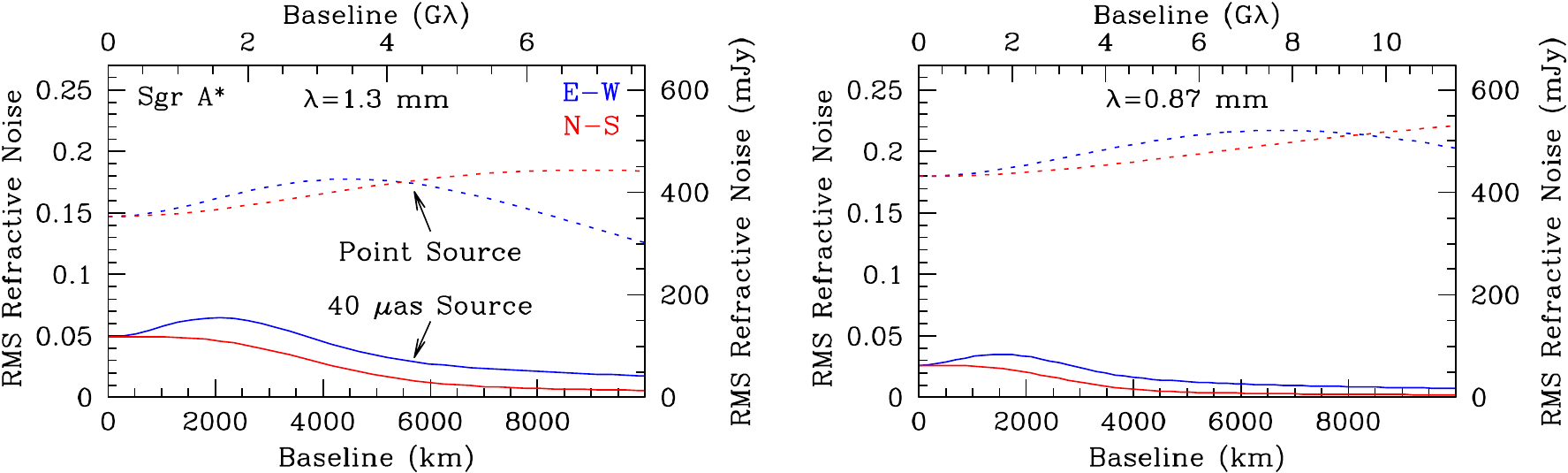}
\caption
{ 
Noise from refractive substructure expected for \sgra\ at $\lambda=1.3~{\rm mm}$ ({\it left}) and at $\lambda=0.87~{\rm mm}$ ({\it right}). For each wavelength, we show the expected noise for both a point source and for a $40~{\mu}{\rm as}$ (FWHM) circular Gaussian source. Estimates given in mJy assume a 2.4 Jy source in both cases. At the shorter wavelength, $\lambda=0.87~{\rm mm}$, the refractive noise for a point source is stronger but is also more heavily quenched by the $40~{\mu}{\rm as}$ source. Note that these estimates do not change sharply with different scattering assumptions, such as a break to Kolmogorov scaling at $\lambda{\sim}1~{\rm cm}$ because the increase in refractive noise from somewhat weaker scattering is partially offset by the concurrent increased quenching from source structure. However, these estimates demonstrate that the refractive noise does depend sensitively on the intrinsic source structure at these wavelengths.
}
\label{fig_EHT_Noise}
\end{figure}

The Event Horizon Telescope (EHT) is an ongoing international effort to establish a global 1.3-mm and 0.87-mm VLBI network \citep{EHT_WhitePaper}. A key motivation for this development is to resolve the intrinsic structure of the Galactic center supermassive black hole, \sgra, which is significantly blurred by scattering at longer wavelengths. \citet{Fish_2014} has shown that remaining blurring in the ensemble-average regime can be effectively removed at $\lambda=1.3~{\rm mm}$. However, while angular broadening decreases steeply with increasing frequency, the fractional effect of refractive noise becomes stronger with increasing frequency, so the influence of refractive substructure remains an important consideration for the EHT. Moreover, because \sgra\ only transitions to weak scattering ($r_0 = r_{\rm F}$) at ${\sim}2\ {\rm THz}$, refractive substructure will affect all foreseeable VLBI of \sgra. 

Indeed, \citet{Gwinn_2014} detected refractive substructure in the scattered image of \sgra\ at $\lambda=1.3~{\rm cm}$, apparent as persistent visibility on baselines that resolved the ensemble-average scattered image. The most pronounced effects of substructure will occur when the angular broadening is comparable to the intrinsic structure (for \sgra, at $\lambda\,{\sim}\,5~{\rm mm}$). Figure~\ref{fig_EHT_Noise} shows the expected refractive noise for \sgra\ at wavelengths and baselines appropriate for the EHT. The noise is significantly affected by the intrinsic source structure, which is currently only well-constrained in the East-West direction \citep{Doeleman_2008,Fish_2011}. With the assumption of a $40\ \mu{\rm as} \approx 4 R_{\rm Sch}$ circular Gaussian source, the refractive noise at $\lambda=1.3~{\rm mm}$ is ${\sim}60~{\rm mJy}$ for long East-West baselines and is ${\sim}25~{\rm mJy}$ for long North-South baselines (the difference is because of ${\sim} 2{:}1$ anisotropy of the scattering). 
Continued observations are essential to better characterize these noise properties. 

Note that while current EHT measurements on long baselines are similar to the level of refractive noise expected for a point source, they cannot be entirely the result of refractive substructure because the ensemble-average visibility would then be far higher than the measured values (i.e., the intrinsic structure would need to be more compact than has been inferred). The consistency of the EHT measurements over observations in different years further argues against refractive noise as a dominant contributor. Nevertheless, because the estimates shown in Figure~\ref{fig_EHT_Noise} anticipate refractive noise at $\lambda=1.3~{\rm mm}$ that is higher than expected thermal noise in EHT measurements \citep[see, e.g.,][]{Lu_2014}, refractive noise will be an important consideration for the interpretation of forthcoming EHT data, especially on baselines for which the ensemble-average visibility is low. Although this noise would remain constant over a single EHT observation and across the bandwidth of the EHT, by 
combining different observations one could reduce the noise and approach the ensemble-average limit \citep{Fish_2014}. 

Refractive image distortion may also cause distortion of the nearly circular photon ring that surrounds the ``shadow'' of the black hole \citep{Falcke_2000}, which may complicate tests of general relativity based on potential asymmetry of the photon ring \citep[e.g.,][]{Johannsen_2010}. While refractive distortion would be stochastic and would not change the shape of the shadow when averaged over many observing nights, it may be an important consideration for single-epoch studies. The supermassive black hole in M87, which subtends a similar angular size but has negligible effects from scattering at EHT wavelengths, will provide an important point of comparison. 

Improved imaging strategies may help to mitigate the effects of refractive noise. However, recall that the correlation length of the noise introduced by substructure is approximately the location at which the ensemble-average visibility falls to $1/\sqrt{e}$ (see \S\ref{sec::A_Visibility}). For a $40~\mu{\rm as}$ circular Gaussian intrinsic source, this length is approximately $2200~{\rm km}$ at $\lambda=1.3~{\rm mm}$ and $1600~{\rm km}$ at $\lambda=0.87~{\rm mm}$. Because the completed EHT will have East-West baselines extending to ${\sim}8000~{\rm km}$ and North-South baselines extending to ${\sim} 11{,}000~{\rm km}$ \citep[see, e.g., Figure 1 in ][]{Fish_2014}, there may be many uncorrelated elements of the noise in a single observation, complicating mitigation strategies.

\section{Summary}
\label{sec::Summary}

Refractive scattering poses a complex observational challenge. It introduces effects that are long-lived and wideband, and that can be difficult to unambiguously identify without multi-epoch observations. 
We have explored the contributions of a particular refractive effect: scattering-induced substructure in images. Building upon the theory and methodology developed by \citetalias{GoodmanNarayan89}, we have shown that small-scale substructure persists in the presence of an extended source. In particular, the action of scattering is \emph{not} a convolution in the average-image regime, which describes nearly all foreseeable VLBI of extended sources. As a result, when the scatter-broadening is comparable to or exceeds the intrinsic structure, the signature consequence of refractive substructure -- elevated visibility on long baselines -- can persist even on interferometric baselines that are long enough to resolve the \emph{unscattered} source. These long-baseline visibilities are not an indication of compact source structure. This result starkly contrasts with the behavior of the diffraction pattern in the observing plane, which is smoothed by an extended source by convolution.

Our results have immediate implications for VLBI studies of galactic nuclei, including studies of AGN with RadioAstron and of \sgra\ with the EHT. Refractive noise is an especially important and subtle consideration for resolved images when the scatter-broadening is comparable or somewhat subdominant to the intrinsic source structure. The refractive noise limits direct imaging to a maximal resolution determined by the scattering and can introduce spurious compact features into images. Nevertheless, our results also indicate that these instruments may provide valuable insight into the turbulent ISM from the new perspective that refractive substructure affords. For instance, substructure reflects properties of the turbulence on much larger scales than those that produce the angular broadening. Also, unlike refractive flux modulation, refractive noise on long baselines does not need to be disentangled from intrinsic source variability; it may, therefore, be a more robust observable than flux 
modulation and an easier measurement than position wander (which requires absolute phase calibration) or image distortion (which requires precise amplitude calibration). Refractive noise can also decouple the intrinsic source structure from scatter broadening without requiring assumptions about the frequency dependence of each \citep[e.g.,][]{Gwinn_2014}.

Ultimately, even refractive noise may not determine a fundamental resolution limit for VLBI. Famously, scattering processes do not destroy information, but rather they add information until the original information is obscured. For instance, in the diffractive regime one can utilize the scattering to \emph{improve} the resolution of an instrument by employing the scattering material as an enormous interstellar lens \citep{Salpeter_1967,Cohen_1967,Lovelace_Thesis,Backer_75,Cordes83,Narayan_1992,CornwellNarayan93,ISO}. Future work based on simplified models for the scattering may suggest superior mitigation and analysis strategies in the refractive regime as well. 

\acknowledgments
We gratefully acknowledge helpful conversations with Ramesh Narayan, Vincent Fish, and Yuri Kovalev. We thank the referee for identifying the connection with geometrical optics. We thank the U.S.\ National Science Foundation (AST-1008865); MJ thanks the Gordon and Betty Moore Foundation (\#GBMF-3561) for financial support for this work.

\begin{appendix}

\section{The Phase Structure Function}
\label{sec::PSF}


To encompass the phase structure function in a simple functional form, we use the following expression \citepalias[][Eq.~3.1.1]{GoodmanNarayan89}: 
\begin{align}
\label{eq::Dphi}
D_{\phi}(\textbf{r}) \approx 
   \begin{cases}
	   C \left[ \left( \left| \textbf{r} \right|^2 + r_{\rm in}^2 \right)^{\alpha/2} - r_{\rm in}^{\alpha} \right], & \text{if } \left| \textbf{r} \right| \ll r_{\rm out}\\
	   C r_{\rm out}^\alpha  & \text{if } \left| \textbf{r} \right| \gg r_{\rm out}.
   \end{cases}
\end{align}
In this expression, $C \propto \nu^{-2}$, reflecting the behavior of the index of refraction for frequencies much greater than the plasma frequency (${\rm kHz}$ in the interstellar medium). Thus, for a pure power-law, the phase coherence length ($D_{\phi}(r_0) \equiv 1$) satisfies $r_0 \propto \nu^{2/\alpha}$. For a negligible inner scale, $r_{\rm in} \ll r_0$, we have
\begin{align}
D_{\phi}(\textbf{r}) \approx \left( \frac{\left| \textbf{r} \right|}{r_0} \right)^\alpha\!.
\end{align}
On the other hand, if $r_{\rm in} \gg r_0$ then
\begin{align}
D_{\phi}(\textbf{r}) \approx 
   \begin{cases}
	   \left( \frac{\left| \textbf{r} \right|}{r_0} \right)^2 & \text{if } r \ll r_{\rm in}, \\
	   \frac{2}{\alpha} \left( \frac{r_{\rm in}}{r_0} \right)^{2-\alpha} \left( \frac{\left| \textbf{r} \right|}{r_0} \right)^\alpha   & \text{if } r \gg r_{\rm in}.
   \end{cases}
\end{align}
While the power-law form at large $r$ is equivalent in the two cases, the correspondence between the $C$ and $r_0$ is affected by an inner scale. This change is what causes the minor modifications to refractive noise on baselines shorter than $\sim \frac{r_0}{r_{\rm in}} r_{\rm R}$ from an inner scale. 

Finally, we account for anisotropic scattering for a pure power law using the form 
\begin{align}
\label{eq::Dphi_Anisotropic}
D_{\phi}(\{ x, y \}) = \left( \frac{x^2}{r_{0,x}^2} +  \frac{y^2}{r_{0,y}^2}\right)^{\alpha/2}\!\!.
\end{align}

\section{Detailed Calculation of Refractive Noise}
\label{sec::Refractive_Noise}

We now calculate the variance of the refractive noise (i.e., the noise $\Delta V_{\rm a} \equiv V_{\rm a} - \left \langle V_{\rm a} \right \rangle_{\rm ea} = V_{\rm a} - V_{\rm ea}$ in the average visibility). This variance is used in \S\ref{sec::Manifestations_of_Refractive_Noise} to quantify various manifestations of refractive noise. Our treatment closely follows \citetalias{GoodmanNarayan89}, although we do not restrict ourselves to long baselines or isotropic scattering, and we include effects from an extended source. 

We first, in \S\ref{sec::Vss}, derive a general expression for the second moment of the snapshot visibility. Then, in \S\ref{sec::RefractiveNoise}, we isolate the contribution of refraction noise and obtain a simplified expression for it in the strong-scattering regime.

\subsection{The Second Moment of $|V_{\rm ss}|$}
\label{sec::Vss}

To calculate the variance of the snapshot visibility modulus, we must multiply the expression for the snapshot visibility (Eq.\ \ref{eq::Vsnapshot}) by its conjugate before evaluating the ensemble average over the screen phases. The ensemble-average over screen phases only affects one term in the resulting integral, which we can readily reduce, as for the ensemble-average visibility (\S\ref{sec::EA_Visibility}): 
\begin{align}
\label{eq::StructureFunctionExp}
\left \langle e^{i \left[\phi(\textbf{x}_1) - \phi(\textbf{x}_2) - \phi(\textbf{x}_3) + \phi(\textbf{x}_4) \right]} \right \rangle_{\rm ea} &= e^{ -\frac{1}{2} \left \langle \left( \phi(\textbf{x}_1) - \phi(\textbf{x}_2) - \phi(\textbf{x}_3) + \phi(\textbf{x}_4) \right)^2 \right \rangle_{\rm ea} }\\
\nonumber &=  e^{-\frac{1}{2} \left[D_\phi(\Delta \textbf{x}_{13}) + D_\phi(\Delta \textbf{x}_{24}) + D_\phi(\Delta \textbf{x}_{12}) - D_\phi(\Delta \textbf{x}_{14}) - D_\phi(\Delta \textbf{x}_{23}) + D_\phi(\Delta \textbf{x}_{34})\right]},
\end{align}
where $\Delta \textbf{x}_{ij} \equiv \textbf{x}_i - \textbf{x}_j$. The second equality follows from
\begin{align}
\left( \phi_1 - \phi_2 - \phi_3 + \phi_4 \right)^2 = (\phi_1 - \phi_3)^2 + (\phi_2 - \phi_4)^2 + (\phi_1 - \phi_2)^2 - (\phi_1 - \phi_4)^2 - (\phi_2 - \phi_3)^2 + (\phi_3 - \phi_4)^2.
\end{align}
Using a Hadamard transform, we then change variables to exploit this pairwise representation:
\begin{align}
  \begin{pmatrix} \textbf{y}_1 \\ \textbf{y}_2 \\ \textbf{y}_3 \\ \textbf{y}_4 \end{pmatrix}
  = \frac{1}{2} \begin{pmatrix} 1 & 1 & 1 & 1 \\ 1 & -1 & 1 & -1 \\ 1 & 1 & -1 & -1 \\ 1 & -1 & -1 & 1 \end{pmatrix}
  \begin{pmatrix} \textbf{x}_1 \\ \textbf{x}_2 \\ \textbf{x}_3 \\ \textbf{x}_4 \end{pmatrix}.
\end{align}
Finally, we can apply the identity $x_1^2 - x_2^2 - x_3^2 + x_4^2 = 2\left( \textbf{y}_1 \cdot \textbf{y}_4 + \textbf{y}_2 \cdot \textbf{y}_3 \right)$ to obtain
\begin{align}
\label{eq::VssStep}
\left \langle \left| V_{\rm ss}(\textbf{b}) \right|^2 \right \rangle_{\rm ea} = &\frac{1}{\left( 2\pi r_{\rm F}^2 \right)^{4} }  \int d^2\textbf{y}_i\, e^{i r_{\rm F}^{-2} \left[ \left( \textbf{y}_1 \cdot \textbf{y}_4 + \textbf{y}_2 \cdot \textbf{y}_3 \right) + (1+M)^{-1} \textbf{b} \cdot \textbf{y}_3 \right] }  \\
\nonumber & {} \times e^{-\frac{1}{2} \left[D_\phi(\Delta \textbf{x}_{13}) + D_\phi(\Delta \textbf{x}_{24}) + D_\phi(\Delta \textbf{x}_{12}) - D_\phi(\Delta \textbf{x}_{14}) - D_\phi(\Delta \textbf{x}_{23}) + D_\phi(\Delta \textbf{x}_{34})\right]}\\
\nonumber & {} \times V_{\rm src}(-\left(1+M\right)\left(\textbf{y}_2 + \textbf{y}_4 \right)) V_{\rm src}^\ast(\left(1+M\right)\left(\textbf{y}_4 - \textbf{y}_2 \right)).
\end{align}
Because the $\Delta \textbf{x}_{ij}$ can be written using only $\{\textbf{y}_{2,3,4} \}$, we can integrate Eq.~\ref{eq::VssStep} over $\textbf{y}_1$ to replace the exponential in $\textbf{y}_1$ by $\left(2\pi r_{\rm F}^2\right)^2 \delta( \textbf{y}_4 )$. We then integrate over $\textbf{y}_4$ to leave
\begin{align}
\label{eq::Vss}
\left \langle |V_{\rm ss}(\textbf{b})|^2 \right \rangle_{\rm ea} 
= \frac{1}{\left( 2\pi r_{\rm F}^2 \right)^2 } \int d^2 \textbf{y}_{2,3}\, e^{i r_{\rm F}^{-2} \left[ \textbf{y}_2 + (1+M)^{-1}\textbf{b} \right] \cdot \textbf{y}_3 }  
& e^{-\frac{1}{2} \left[2 D_\phi(\textbf{y}_{3}) + 2 D_\phi(\textbf{y}_{2}) - D_\phi(\textbf{y}_{2} + \textbf{y}_3) - D_\phi(\textbf{y}_{2} - \textbf{y}_3) \right]} \left| V_{\rm src}\left( (1+M)\textbf{y}_2 \right) \right|^2.
\end{align}
Note that the phase structure function is still completely general, as is the intrinsic source structure. Moreover, we have not yet made any approximations based on the strength of the phase fluctuations.

\subsection{Refractive Noise}
\label{sec::RefractiveNoise}

\begin{figure}[t]
\centering
\includegraphics[width=0.7\textwidth]{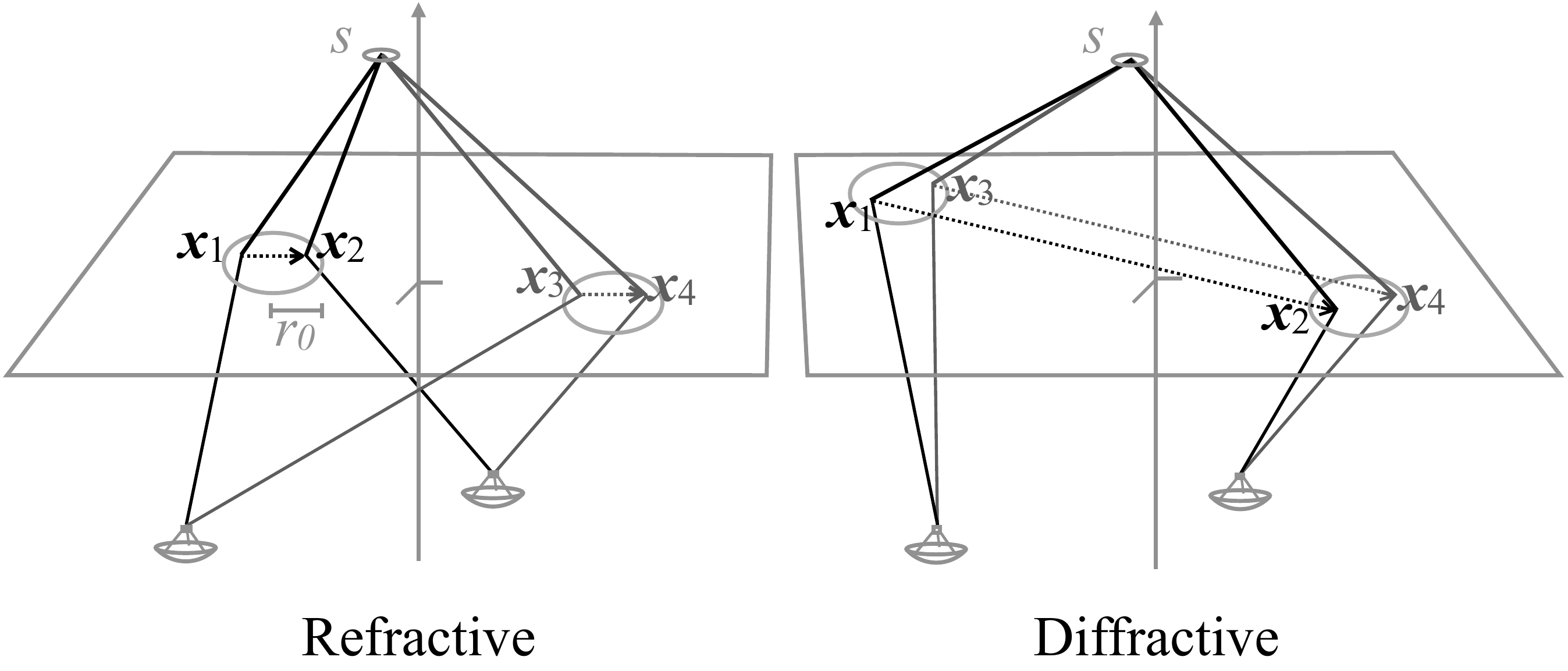}
\caption
{ 
Refractive and diffractive contributions to the second moment of the snapshot visibility.
}
\label{fig_noise_geometry}
\end{figure}

Three effects contribute to a snapshot visibility: diffractive noise, refractive noise, and $V_{\rm ea}(b)$. However, for the second moment, as expressed in Eq.~\ref{eq::Vss}, each arises from a different region of the integral. In the strong scattering regime with $\alpha < 2$, the dominant contributions to the integral have at least one of $\{ \left| \textbf{y}_2 \right|, \left| \textbf{y}_3 \right| \} \lsim r_0$. The region $\{\left| \textbf{y}_2 \right| \gg r_0,\,\left| \textbf{y}_3 \right| \lsim r_0\}$ gives the diffractive noise, as is evident by its suppression from an extended source. The remaining region of the integral gives the second moment of $|V_{\rm a}|$. Figure \ref{fig_noise_geometry} illustrates this geometry.

To isolate the noise $\Delta V_{\rm a}$, note that the contribution of $V_{\rm ea}$ can be written as 
\begin{align}
\label{eq::VeaNoise}
\frac{1}{\left( 2\pi r_{\rm F}^2 \right)^2 } \int d^2 \textbf{y}_{2,3}\, e^{i r_{\rm F}^{-2} \left[ \textbf{y}_2 + (1+M)^{-1}\textbf{b} \right] \cdot \textbf{y}_3 } e^{-D_{\phi}(\textbf{y}_2)} \left| V_{\rm src}\left( (1+M)\textbf{y}_2 \right) \right|^2 = e^{-D_{\phi}((1+M)^{-1}\textbf{b})} \left| V_{\rm src}\left( \mathbf{b} \right) \right|^2 = \left| V_{\rm ea}(\mathbf{b}) \right|^2.
\end{align}

Because they focused on long baseline properties, \citepalias{GoodmanNarayan89} neglected the phase variations $e^{i r_{\rm F}^{-2} \textbf{y}_2 \cdot \textbf{y}_3 }$ in Eq.\ \ref{eq::Vss}, which are then small over the region that dominates the refractive contribution, giving an $\mathcal{O}((1+M)r_0/\textbf{b})^2$ correction. With this approximation, and in the strong scattering regime, the integrals in Eq.\ \ref{eq::VeaNoise} factorize \citep{Johnson_Thesis}, simplifying the remaining calculation. However, we must retain the coupling term in order to understand the refractive noise on all baselines. 

The integral over $\textbf{y}_3$ is the one that is challenging numerically, but we can get it into a manageable form. Because $D_{\phi}(\textbf{y}) = D_{\phi}(-\textbf{y})$ and $|\textbf{y}_3| \gg |\textbf{y}_2|$ for the average visibility, we have that
\begin{align}
\label{eq::Dphi_Expand}
2 D_\phi(\textbf{y}_{3}) - D_\phi(\textbf{y}_{2} + \textbf{y}_3) - D_\phi(\textbf{y}_{2} - \textbf{y}_3) \approx \left. -\left( \textbf{y}_2 \cdot \nabla \right)^2 D_{\phi}(\textbf{y}) \right\rfloor_{\textbf{y} = \textbf{y}_3}\!.
\end{align}
Importantly, this representation does not assume that $D_{\phi}(\textbf{y})$ is isotropic. 

Combining Eq.~\ref{eq::Vss}, Eq.~\ref{eq::VeaNoise}, and Eq.~\ref{eq::Dphi_Expand} then gives
\begin{align}
\label{2D_general}
\nonumber \left \langle |\Delta V_{\rm a}(\textbf{b})|^2 \right \rangle_{\rm ea} &= \nonumber \left \langle |V_{\rm a}(\textbf{b})|^2 \right \rangle_{\rm ea} - \left|V_{\rm ea}\right|^2\\
\nonumber &= \frac{1}{\left( 2\pi r_{\rm F}^2 \right)^2 } \int d^2 \textbf{y}_{2,3}\, e^{\frac{i}{r_{\rm F}^{2}} \left[ \textbf{y}_2 + \frac{\textbf{b}}{1+M} \right] \cdot \textbf{y}_3 }  
\left[ e^{-\frac{1}{2} \left[2 D_\phi(\textbf{y}_{3}) - D_\phi(\textbf{y}_{2} + \textbf{y}_3) - D_\phi(\textbf{y}_{2} - \textbf{y}_3) \right]}  - 1 \right] e^{-D_{\phi}(\textbf{y}_2)} \left| V_{\rm src}\left( (1+M)\textbf{y}_2 \right) \right|^2\\
\nonumber &\approx \frac{1}{\left( 2\pi r_{\rm F}^2 \right)^2 } \int d^2 \textbf{y}_{2,3}\, e^{\frac{i}{r_{\rm F}^{2}} \left[ \textbf{y}_2 +  \frac{\textbf{b}}{1+M} \right] \cdot \textbf{y}_3 } \left[ \left. \frac{1}{2} \left( \textbf{y}_2 \cdot \nabla \right)^2 D_{\phi}(\textbf{y}) \right\rfloor_{\textbf{y} = \textbf{y}_3} \right] \left| V_{\rm ea}\left( (1+M)\textbf{y}_2 \right) \right|^2\\
\nonumber &\approx \frac{-1}{\left( 2\pi r_{\rm F}^2 \right)^2 } \int d^2 \textbf{y}_{2,3}\, \frac{1}{2 r_{\rm F}^4} \left( \textbf{y}_2 \cdot  \left[ \textbf{y}_2 + (1+M)^{-1}\textbf{b} \right] \right)^2  e^{\frac{i}{r_{\rm F}^{2}} \left[ \textbf{y}_2 + (1+M)^{-1}\textbf{b} \right] \cdot \textbf{y}_3 } D_{\phi}(\textbf{y}_3) \left| V_{\rm ea}\left( (1+M)\textbf{y}_2 \right) \right|^2\\
&\equiv \frac{-1}{\left( 2\pi r_{\rm F}^2 \right)^2 } \int d^2 \textbf{y}_{2}\, \frac{1}{2 r_{\rm F}^4} \left( \textbf{y}_2 \cdot  \left[ \textbf{y}_2 + (1+M)^{-1}\textbf{b} \right] \right)^2 \tilde{D}_{\phi}\left(\textbf{y}_2 + (1+M)^{-1}\textbf{b}\right) \left| V_{\rm ea}\left( (1+M)\textbf{y}_2 \right) \right|^2,
\end{align}
where we have introduced
\begin{align}
\tilde{D}_{\phi}(\textbf{y}) \equiv \int d^2 \textbf{y}_3 e^{\frac{i}{r_{\rm F}^{2}} \textbf{y} \cdot \textbf{y}_3 }  D_{\phi}(\textbf{y}_3).
\end{align}
Once $\tilde{D}_{\phi}(\textbf{y})$ is specified (analytically), the integral in Eq.\ \ref{2D_general} can easily be performed numerically. 

Note that $\tilde{D}_{\phi}(\textbf{y}_3)$ is closely related to the power spectrum of phase fluctuations at a scale $\textbf{q} \equiv \textbf{y}_3/r_{\rm F}^2$ \citep[see, e.g.,][]{Tatarskii_1971,Goodman_Narayan_1985,Coles_1987,Lambert_Rickett_2000}. Specifically, after removing a constant phase offset, which does not affect the snapshot visibilities, $\tilde{D}_{\phi}\left( \textbf{y}_3 \right) \propto -Q\left( \textbf{q} \equiv \textbf{y}_3/r_{\rm F}^2 \right)$ . Hence, positivity of the power spectrum enforces positivity of the refractive noise. For instance, consider an isotropic power-law structure function: $D_{\phi}(\textbf{y}_3) = \left| \frac{\textbf{y}_3}{r_{0}}\right|^{\alpha}$. Then
\begin{align}
\label{eq::Q_isotropic}
\tilde{D}_{\phi}\left(\textbf{y}_3 \right) = -2^{1 + \alpha} \pi \alpha \frac{\Gamma(1 + \alpha/2)}{\Gamma(1-\alpha/2)} r_0^{-\alpha} \left| \frac{\textbf{y}_3}{r_{\rm F}^2}\right|^{-(2 + \alpha)}\!\!.
\end{align}
We can easily generalize this result to include inner and outer scales in $D_{\phi}(\textbf{x})$ or anisotropy. For instance, 
\begin{align}
\label{eq::Q_general}
\tilde{D}_{\phi}\left(\textbf{y}_3 \right) \sim -2^{2 + \alpha} \pi \frac{\Gamma(1 + \alpha/2)}{\Gamma(-\alpha/2)} r_0^{-\alpha} \left[ \left| \frac{\textbf{y}_3}{r_{\rm F}^2}\right|^2 + \left( \frac{1}{ r_{\rm out}}\right)^2 \right]^{-(1 + \alpha/2)} e^{-\left|\frac{r_{\rm in}}{r_0} \frac{\textbf{y}_3 }{r_{\rm R}} \right|^2}.
\end{align}
Similarly, to account for anisotropic scattering, we use the form for $D_{\phi}(\{ x, y \})$ given in Eq.~\ref{eq::Dphi_Anisotropic}. Then
\begin{align}
\label{Dtile_Anisotropic}
\tilde{D}_{\phi}(\textbf{y}) = -2^{2 + \alpha} \pi \frac{\Gamma(1 + \alpha/2)}{\Gamma(-\alpha/2)} r_{\rm F}^{4 + 2\alpha} r_{0,x} r_{0,y} \left( r_{0,x}^2 x^2 + r_{0,y}^2 y^2 \right)^{-(1 + \alpha/2)}.
\end{align}

\section{Scattering Simulations}
\label{sec::Simulations}

In this section, we provide additional details about the scattering simulations shown in Figure~\ref{fig_image_examples}. Similar simulations have been performed in many contexts \citep[e.g.,][]{Coles_1995,Coles_2010,Habibi_2013}, but ours are perhaps most similar to the one-dimensional simulations of \citetalias{NarayanGoodman89}, with the primary difference that we are simulating a two-dimensional scattering screen.  

We first generate an $N{\times}N$ grid of independent zero-mean, complex Gaussian random variables. Next, we imprint the appropriate power spectrum $Q$ of phase fluctuations by 
taking the discrete Fourier transform of the phases, multiplying by $\sqrt{Q(\textbf{q})}$, and then inverse Fourier transforming the result. Finally, we calculate the empirical structure function and apply an overall normalization to the phases so that $D_{\phi}(r_0) = 1$. 

For the power spectrum, we employ the commonly used form
\begin{align}
\label{eq::spectrum}
Q(\textbf{q}) \propto \left( |\textbf{q}|^2 + q_{\rm min}^2 \right)^{-(1+\alpha/2)} e^{-\left( \frac{|\textbf{q}|}{q_{\rm max}} \right)^2},
\end{align}
which corresponds to a phase structure function with an index $\alpha$, inner scale $r_{\rm in} \sim 1 / q_{\rm max}$, and outer scale $r_{\rm out} \sim 1 / q_{\rm min}$. Note that specific conventions for the form of Eq.~\ref{eq::spectrum} and the relationship between $r_{\rm in/out}$ and $q_{\rm max/min}$ vary throughout the literature. 

After calculating the screen phases, we can evaluate the electric field in the observing plane (Eq.~\ref{eq::field}), snapshot visibilities (Eq.~\ref{eq::Vsnapshot}), or the scattered image (Eq.~\ref{eq::Image_ss}). For the scattered images shown in Figure \ref{fig_image_examples}, we calculated the phase gradient at each location using a discrete approximation with nearest-neighbor grid points and then utilized the approximation of Eq.~\ref{eq::Image_Approx}.

The greatest computational hurdle is then storing the random phase screen in memory; the number of screen phases that must be stored is $\sim (r_{\rm R}/r_0)^2 = (r_{\rm F}/r_0)^4$. To achieve the largest ratios $r_{\rm F}/r_0 \gg 1$, appropriate for strong scattering, we set $r_0$ to be equal to the grid spacing. Our simulations have a linear size of $2^{14}$ points, so we can readily achieve $r_{\rm F}/r_0 \sim 35$. However, because the phase varies linearly on scales shorter than $r_{\rm in}$, one only needs to resolve the larger of $\{ r_0,\, r_{\rm in} \}$. Hence, simulations could be performed with a much larger $r_{\rm F}/r_0$ if they also included a large inner scale.

\end{appendix}

\ \\

\bibliography{Refractive_Substructure.bib}

\end{document}